\def\year{2020}\relax
\documentclass[letterpaper]{article} 
\usepackage{aaai20}  
\usepackage{times}  
\usepackage{helvet} 
\usepackage{courier}  
\usepackage[hyphens]{url}  
\usepackage{graphicx} 
\usepackage{graphicx}  
\usepackage{graphicx, subfigure}
\usepackage{amsmath}
\usepackage{amssymb}
\usepackage{algorithm}
\usepackage{algorithmic}
\usepackage{array}
\usepackage{fixltx2e}
\usepackage{dblfloatfix}
\usepackage{mathrsfs}
\usepackage{multirow}
\usepackage{subfigure}
\title{A Generalized Framework for Edge-preserving and Structure-preserving Image Smoothing}
\author{Wei Liu$^{1,2}$, Pingping Zhang$^{3}$, Yinjie Lei$^{4\ast}$, Xiaolin Huang$^{1,5}$, Jie Yang$^{1,5}$\thanks{Jie Yang and Yinjie Lei are the corresponding authors of this paper.}, Ian Reid$^{2}$\\
$^1$Department of Automation, Shanghai Jiao Tong University, $^2$The University of Adelaide\\
$^3$Dalian University of Technology, $^4$Sichuan University, $^5$Institute of Medical Robotics, Shanghai Jiao Tong University\\
{\small {\{wei.liu02, ian.reid\}@adelaide.edu.au,  jssxzhpp@mail.dlut.edu.cn,  yinjie@scu.edu.cn,  \{xiaolinhuang, jieyang\}@sjtu.edu.cn}}
}

\begin{document}

\maketitle

\begin{abstract}
Image smoothing is a fundamental procedure in applications of both computer vision and graphics. The required smoothing properties can be different or even contradictive among different tasks. Nevertheless, the inherent smoothing nature of one smoothing operator is usually fixed and thus cannot meet the various requirements of different applications. In this paper, a non-convex non-smooth optimization framework is proposed to achieve diverse smoothing natures where even contradictive smoothing behaviors can be achieved. To this end, we first introduce the truncated Huber penalty function which has seldom been used in image smoothing. A robust framework is then proposed. When combined with the strong flexibility of the truncated Huber penalty function, our framework is capable of a range of applications and can outperform the state-of-the-art approaches in several tasks. In addition, an efficient numerical solution is provided and its convergence is theoretically guaranteed even the optimization framework is non-convex and non-smooth. The effectiveness and superior performance of our approach are validated through comprehensive experimental results in a range of applications.
\end{abstract}


\section{Introduction}
\label{SecIntroduction}

The key challenge of many tasks in both computer vision and graphics can be attributed to image smoothing. At the same time, the required smoothing properties can vary dramatically for different tasks. In this paper, depending on the required smoothing properties, we roughly classify a large number of applications into four groups.

Applications in the first group require the smoothing operator to smooth out small details while preserving strong edges, and the amplitudes of these strong edges can be reduced but the edges should be neither blurred nor sharpened. Representatives in this group are image detail enhancement and HDR tone mapping \cite{farbman2008edge,fattal2007multiscale,he2013guided}. Blurring edges can result in halos while sharpening edges will lead to gradient reversals \cite{farbman2008edge}.

\begin{figure}
\subfigure[]
{
\includegraphics[width=0.222\linewidth]{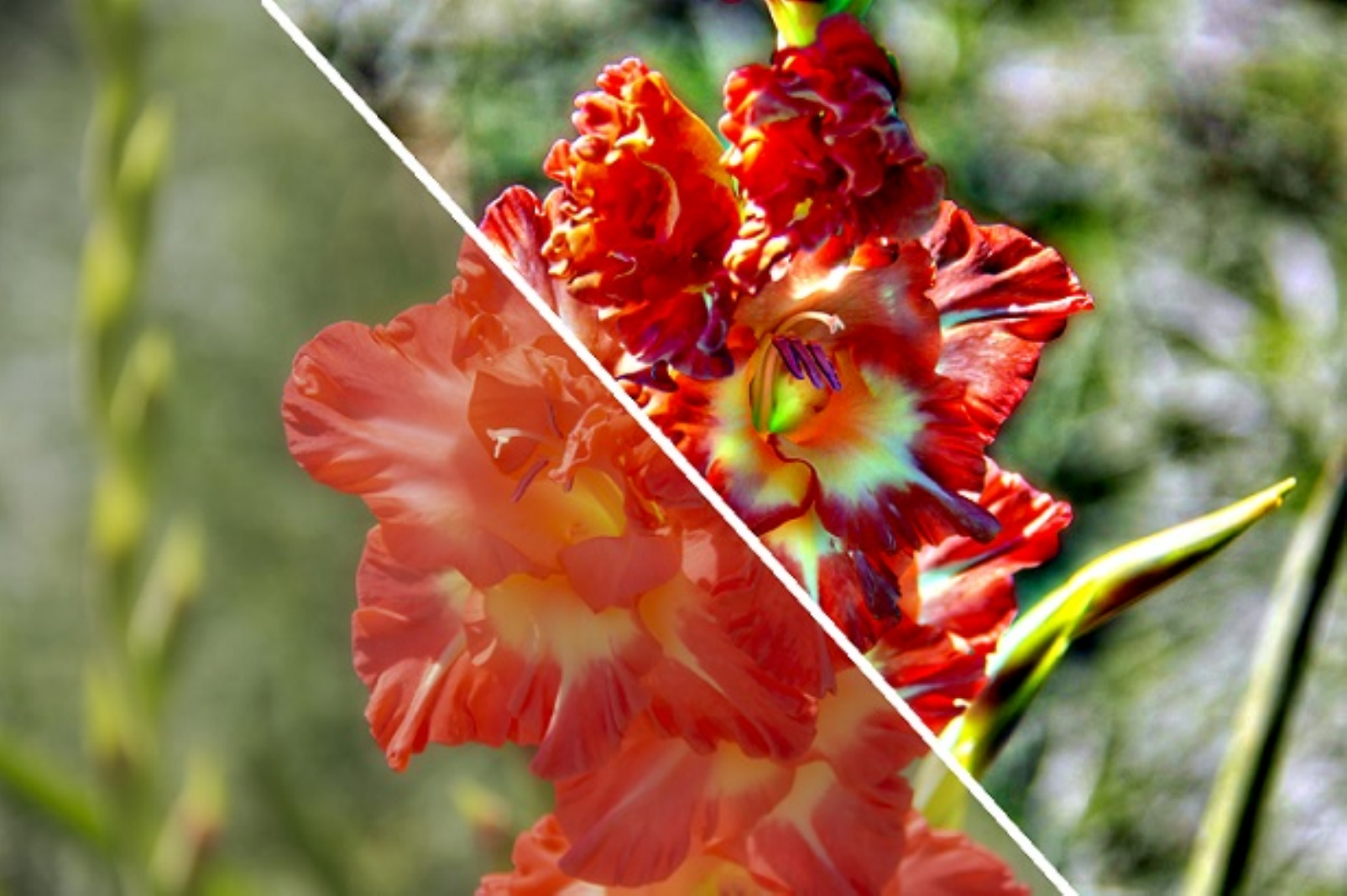}
}
\subfigure[]
{
\includegraphics[width=0.222\linewidth]{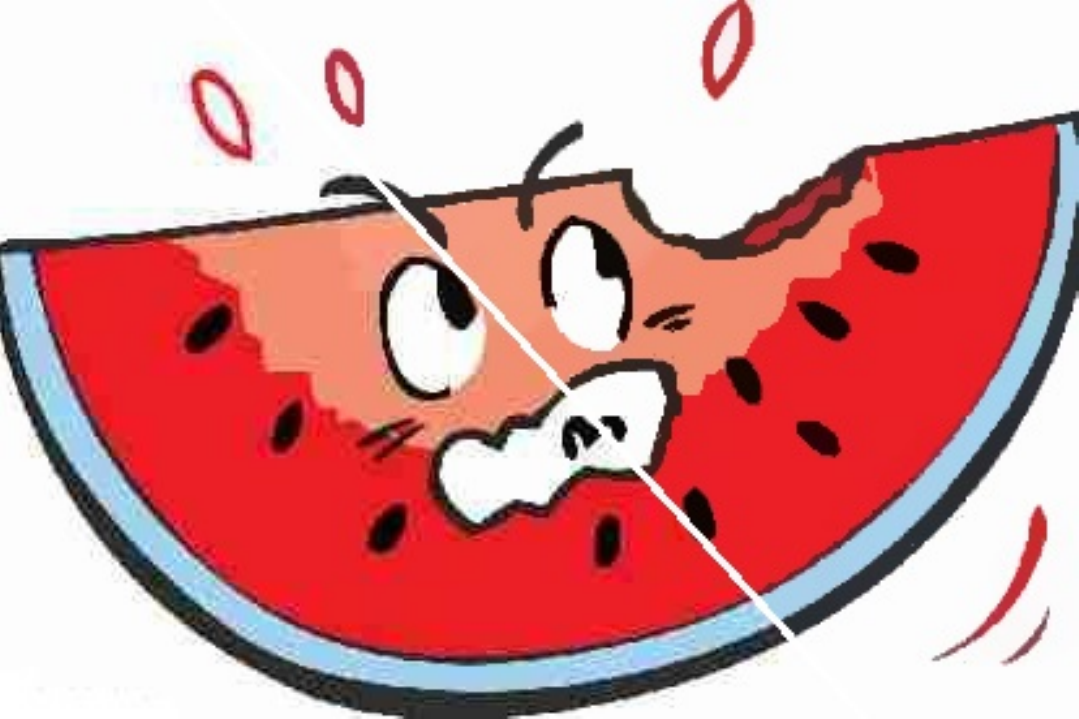}
}
\subfigure[]
{
\includegraphics[width=0.222\linewidth]{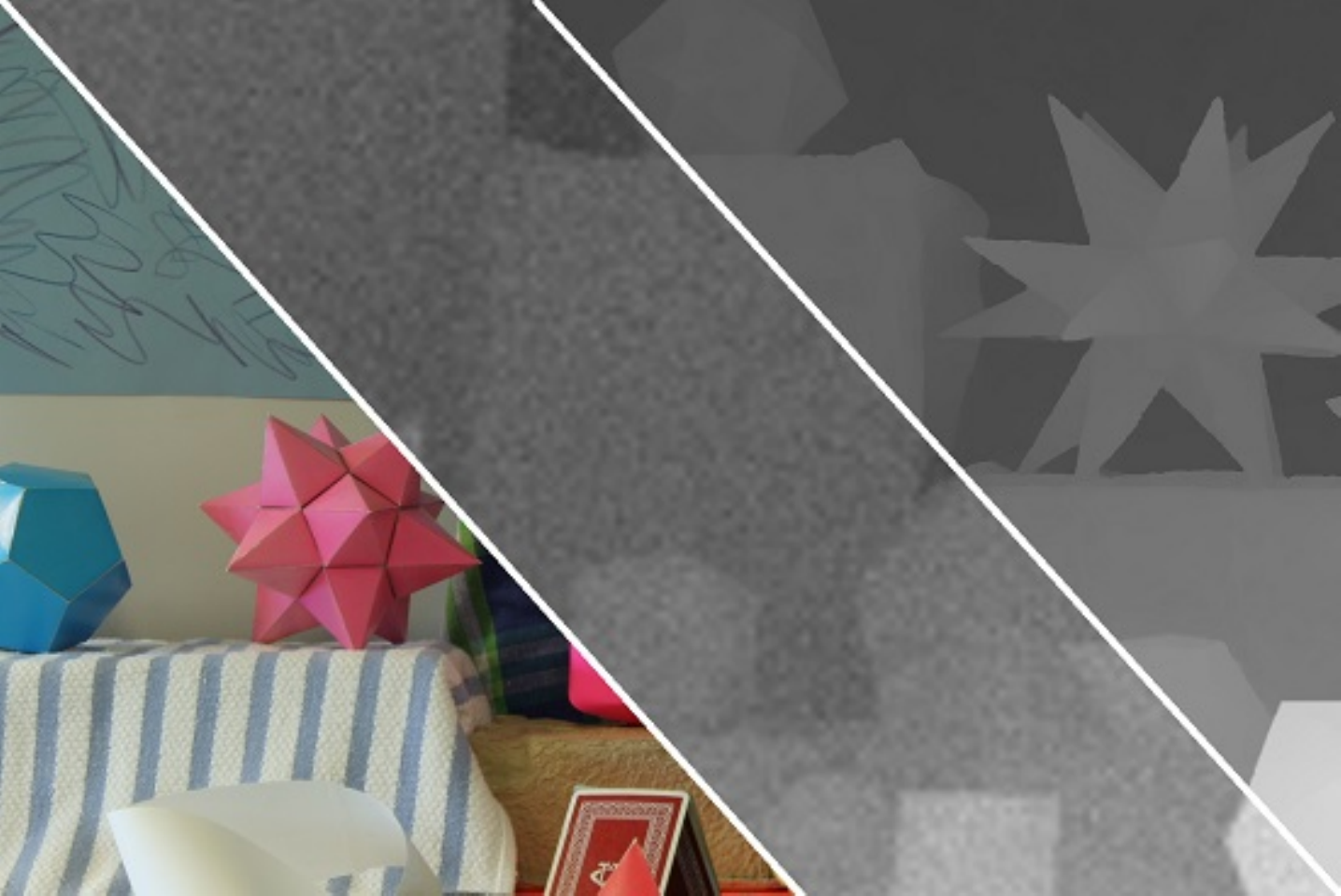}
}
\subfigure[]
{
\includegraphics[width=0.222\linewidth]{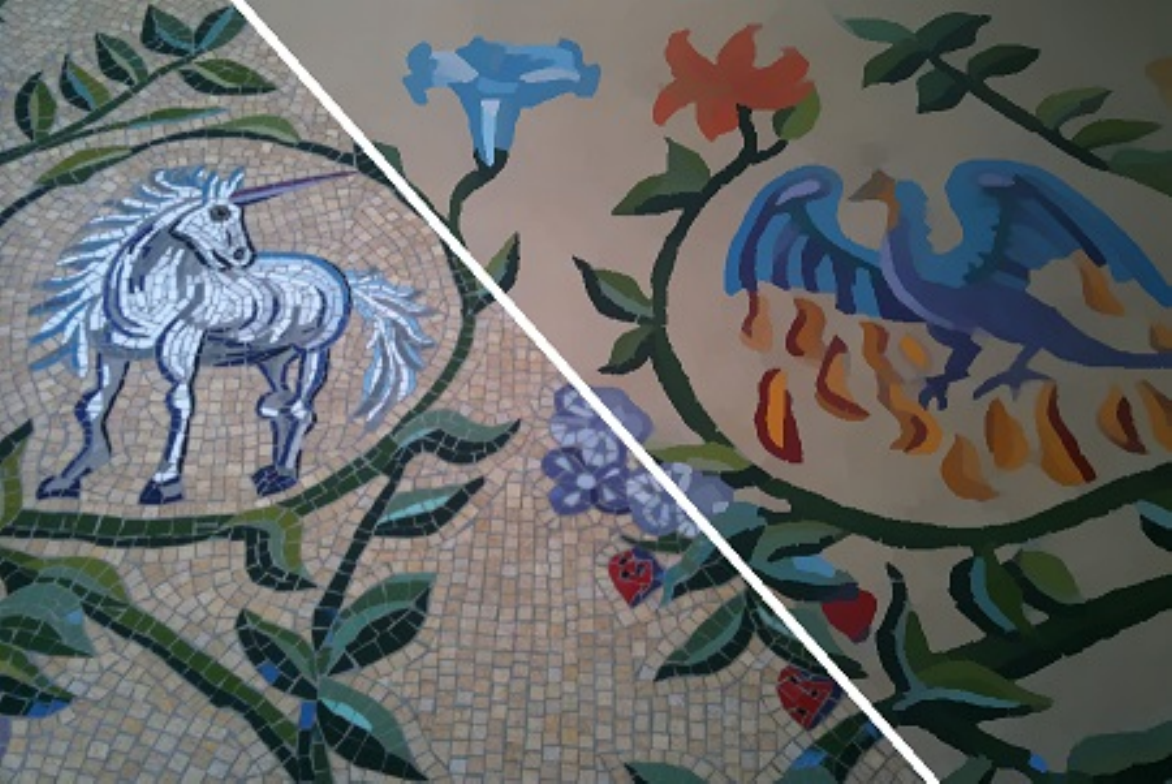}
}
  \caption{Our method is capable of (a) image detail enhancement, (b) clip-art compression artifacts removal, (c) guided depth map upsampling and (d) image texture removal. These applications are representatives of edge-preserving and structure-preserving image smoothing and require contradictive smoothing properties.}\label{FigCover}
\end{figure}

The second group includes tasks like clip-art compression artifacts removal \cite{nguyen2015fast,xu2011image}, image abstraction and pencil sketch production \cite{xu2011image}. In contrast to the ones in the first group, these tasks require to smooth out small details while sharpening strong edges. This is because edges can be blurred in the compressed clip-art image and they need to be sharpened when the image is recovered (see Fig.~\ref{FigCover}(b) for example). Sharper edges can produce better visual quality in image abstraction and pencil sketch. At the same time, the amplitudes of strong edges are not allowed to be reduced in these tasks.

Guided image filtering, such as guided depth map upsampling \cite{park2011high,ferstl2013image,liu2017robust} and flash/no flash filtering \cite{kopf2007joint,petschnigg2004digital}, is categorized into the third group. The structure inconsistency between the guidance image and target image, which can cause blurring edges and texture copy artifacts in the smoothed image \cite{ham2015robust,liu2017robust}, should be properly handled by the specially designed smoothing operator. They also need to sharpen edges in the smoothed image due to the reason that low-quality capture of depth and noise in the no flash images can lead to blurred edge (see Fig.~\ref{FigCover}(c) for example).

Tasks in the fourth group require to smooth the image in a scale-aware manner, e.g., image texture removal \cite{xu2012structure,zhang2014rolling,cho2014bilateral}. This kind of tasks require to smooth out small structures even when they contain strong edges, while large structure should be properly preserved even the edges are weak (see Fig.~\ref{FigCover}(d) for example). This is totally different from that in the above three groups where they all aim at preserving strong edges.

To be more explicit, we categorize the smoothing procedures in the first to the third groups as \emph{edge-preserving image smoothing} since they try to preserve salient edges, while the smoothing processes in the fourth group are classified as \emph{structure-preserving image smoothing} because they aim at preserving salient structures.

A diversity of edge-preserving and structure-preserving smoothing operators have been proposed for various tasks. Generally, each of them is designed to meet the requirements of certain applications, and thus its inherent smoothing nature is usually fixed. Therefore, there is seldom a smoothing operator that can meet all the smoothing requirements of the above four groups, which are quite different or even contradictive. For example, the $L_0$ norm smoothing \cite{xu2011image} can sharpen strong edges and is suitable for clip-art compression artifacts removal, however, this will lead to gradient reversals in image detail enhancement and HDR tone mapping. The weighted least squares (WLS) smoothing \cite{farbman2008edge} performs well in image detail enhancement and HDR tone mapping, but it is not capable of sharpening edges and structure-preserving smoothing, etc.

In contrast to most of the smoothing operators in the literature, a new smoothing operator, which is based on a non-convex non-smooth optimization framework, is proposed in this paper. It can achieve different and even contradictive smoothing behaviors and is able to handle the applications in the four groups mentioned above. The main contributions of this paper are as follows:

\begin{itemize}
  \item[1.] We introduce the \emph{truncated Huber penalty} function which has seldom been used in image smoothing. By varying the parameters, it shows strong flexibility.

  \item[2.] A robust non-convex non-smooth optimization framework is proposed. When combined with the strong flexibility of the truncated Huber penalty function, our model can achieve various and even contradictive smoothing behaviors. We show that it is able to handle the tasks in the four groups mentioned above. This has seldom been achieved by previous smoothing operators.

  \item[3.] An efficient numerical solution to the proposed optimization framework is provided. Its convergence is theoretically guaranteed.

  \item[4.] Our method is able to outperform the specially designed approaches in many tasks and state-of-the-art performance is achieved.
\end{itemize}

\section{Related Work}
\label{SecRelatedWork}

Tremendous smoothing operators have been proposed in recent decades. In terms of edge-preserving smoothing, bilateral filter (BLF) \cite{tomasi1998bilateral} is the early work that has been used in various tasks such as image detail enhancement \cite{fattal2007multiscale}, HDR tone mapping \cite{durand2002fast}, etc. However, it is prone to produce results with gradient reversals and halos \cite{farbman2008edge}. Its alternatives \cite{gastal2012adaptive,gastal2011domain} also share a similar problem. Guided filter (GF) \cite{he2013guided} can produce results free of gradient reversals but halos still exist. The WLS smoothing \cite{farbman2008edge} solves a global optimization problem and performs well in handling these artifacts. The $L_0$ norm smoothing is able to eliminate low-amplitude structures while sharpening strong edges, which can be applied to the tasks in the second group. To handle the structure inconsistency problem, Shen et~al. \cite{shen2015mutual} proposed to perform mutual-structure joint filtering. They also explored the relation between the guidance image and target image via optimizing a scale map \cite{shen2015multispectral}, however, additional processing was adopted for structure inconsistency handling. Ham et~al. \cite{ham2015robust} proposed to handle the structure inconsistency by combining a static guidance weight with a Welsch's penalty \cite{holland1977robust} regularized smoothness term, which leaded to a static/dynamic (SD) filter. Gu et~al. \cite{gu2017learning} presented a weighted analysis representation model for guided depth map enhancement. 

In terms of structure-preserving smoothing, Zhang et~al. \cite{zhang2014rolling} proposed to smooth structures of different scales with a rolling guidance filter (RGF). Cho et~al. \cite{cho2014bilateral} modified the original BLF with local patch-based analysis of texture features and obtained a bilateral texture filter (BTF) for image texture removal. Karacan et~al. \cite{karacan2013structure} proposed to smooth image textures by making use of region covariances that captured local structure and textural information. Xu et~al. \cite{xu2012structure} adopted the relative total variation (RTV) as a prior to regularize the texture smoothing procedure. Fan et~al. \cite{fan2018image,fan2019general} proposed to perform various kinds of image smoothing through convolutional neural networks. Chen et~al. \cite{chan2005aspects} proved that the TV-$L_1$ model \cite{chan2005aspects,nikolova2004variational} could smooth images in a scale-aware manner, and it is thus ideal for structure-preserving smoothing such as image texture removal \cite{buades2010fast}.

Most of the approaches mentioned above are limited to a few applications because their inherent smoothing natures are usually fixed. In contrast, our method proposed in this paper can have strong flexibility in achieving various smoothing behaviors, which enables wider applications of our method than most of them. Moreover, our method can show better performance than these methods in several applications that they are specially designed for.

\section{Our Approach}

\subsection{Truncated Huber Penalty Function}
\label{SecTruncatedHuber}

\begin{figure}
\centering
\subfigure[]
{
\includegraphics[width=0.35\linewidth]{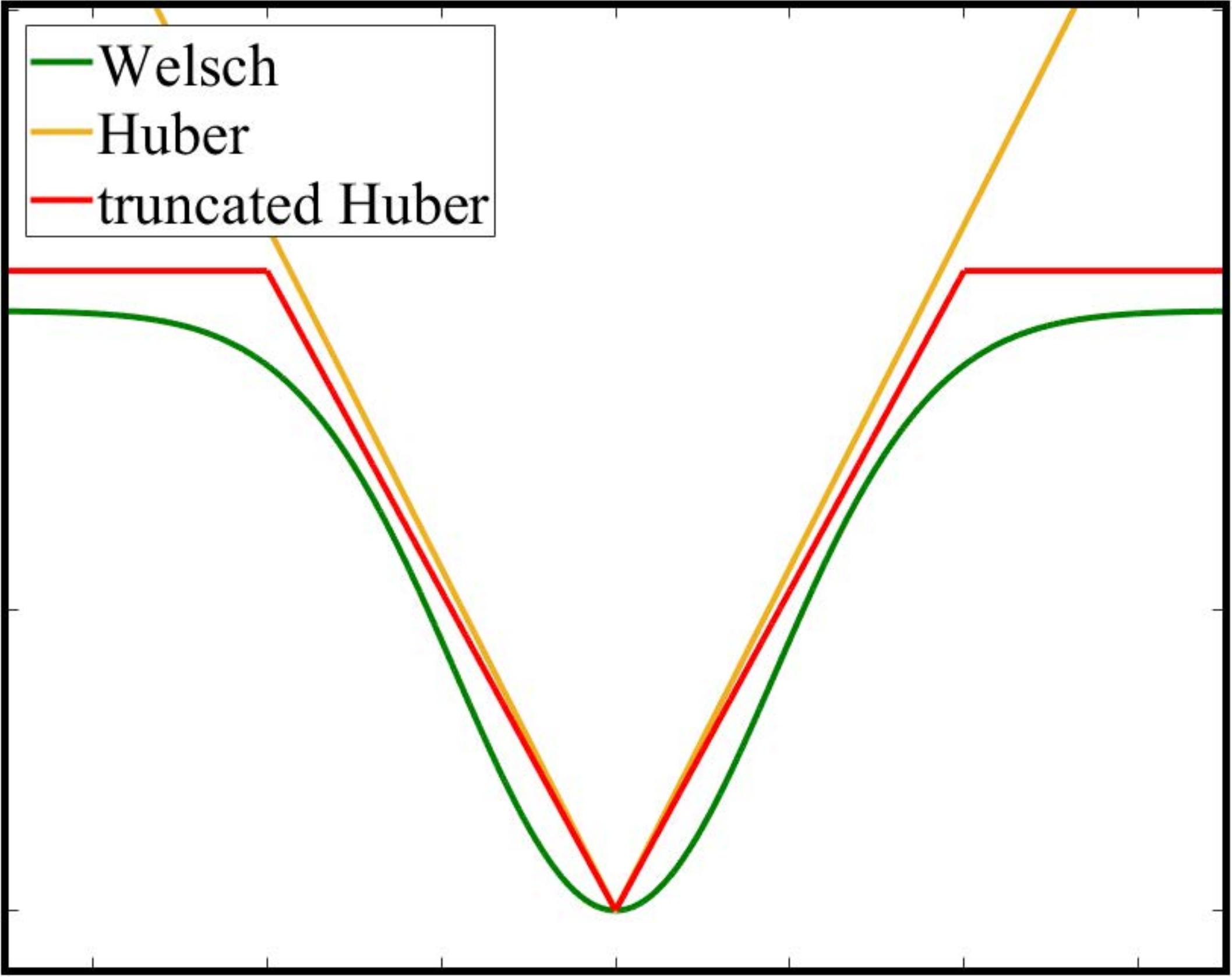}
}
\subfigure[]
{
\includegraphics[width=0.35\linewidth]{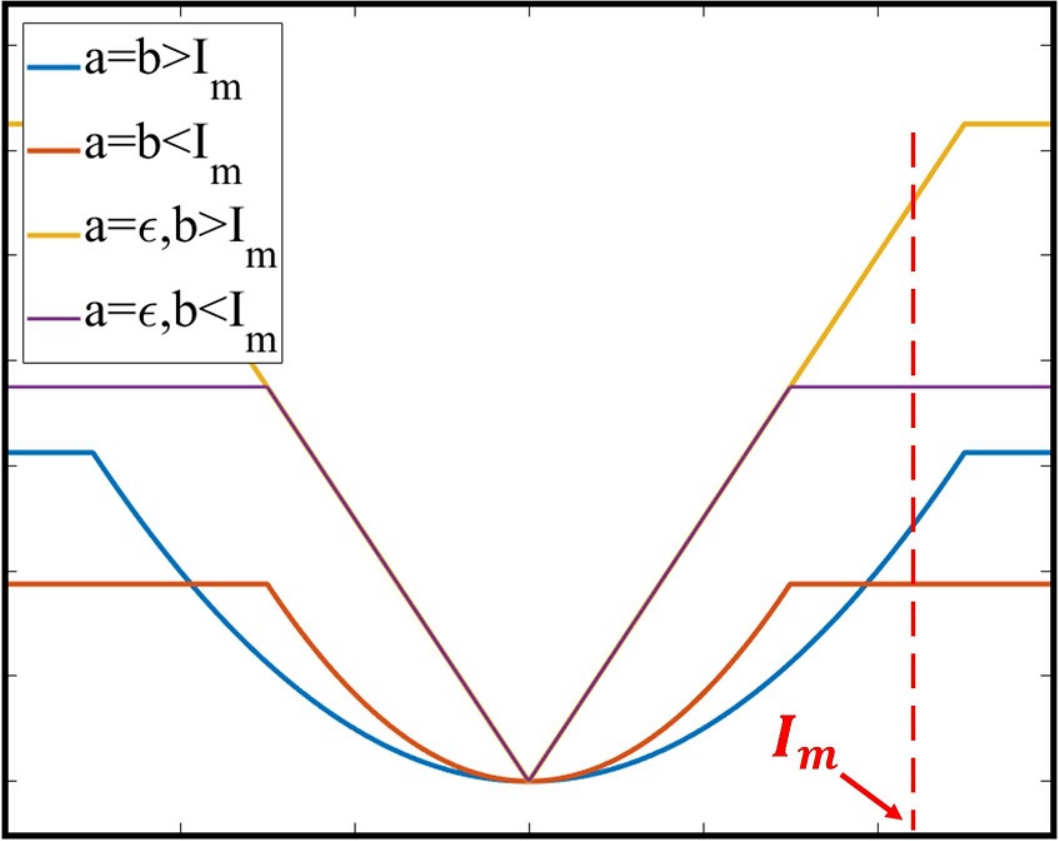}
}
\caption{Plots of (a) different penalty functions and (b) the truncated Huber penalty function with different parameter settings.}\label{FigPenaltyComp}
\end{figure}

We first introduce the truncated Huber penalty function which is defined as:
\begin{small}
\begin{eqnarray}\label{EqTruncatedHuber}
{ h_T(x)=\left\{\begin{array}{l}
   h(x), \ \ \ \ \ |x|\leq b\\
   b -\frac{a}{2}, \ \ \ |x|>b
  \end{array}\right.
  \text{s.t.} \ \ \ a \leq b,
}
\end{eqnarray}
\end{small}
where $a,b$ are constants. $h(\cdot)$ is the Huber penalty function \cite{huber1964robust} defined as:
\begin{small}
\begin{eqnarray}\label{EqHuber}
{
  h(x)=\left\{\begin{array}{l}
   \frac{1}{2a}x^2, \ \ \ \ \ \ |x|<a\\
   |x|-\frac{a}{2}, \ \ |x|\geq a
  \end{array}\right.,
}
\end{eqnarray}
\end{small}
$h_T(\cdot)$ and $h(\cdot)$ are plotted in Fig.~\ref{FigPenaltyComp}(a) with $a=\epsilon$ which is a sufficient small value (e.g., $\epsilon=10^{-7}$). $h(\cdot)$ is an edge-preserving penalty function, but it cannot sharpen edges when adopted to regularize the smoothing procedure. In contrast, $h_T(\cdot)$ can sharpen edges because it is able to not penalize image edges due to the truncation. The Welsch's penalty function \cite{holland1977robust}, which was adopted in the recent proposed SD filter \cite{ham2015robust}, is also plotted in the figure. This penalty function is known to be capable of sharpening edges, which is also because it seldom penalizes strong image edges. The Welsch's penalty function is close to the $L_2$ norm when the input is small, while the $h_T(\cdot)$ can be close to the $L_1$ norm when $a$ is set sufficient small, which demonstrates $h_T(\cdot)$ can better preserve weak edges than the Welsch's penalty function.

With different parameter settings, $h_T(\cdot)$ can show strong flexibility to yield different penalty behaviors. Assume the input intensity values are within $[0, I_m]$, then the amplitude of any edge will fall in $[0, I_m]$. We first set $a=\epsilon$. Then if we set $b>I_m$, $h_T(\cdot)$ will be actually the same as $h(\cdot)$ because the second condition in Eq.~(\ref{EqTruncatedHuber}) can never be met. Because $a$ is sufficient small, $h_T(\cdot)$ will be close to the $L_1$ norm in this case, and thus it will be an edge-preserving penalty function that does not sharpen edges. Conversely, when we set $b<I_m$, the truncation in $h_T(\cdot)$ will be activated. This can lead to having penalization on weak edges without penalizing strong edges, and thus the strong edges are sharpened. To be short, $b$ can act as a switch to decide whether $h_T(\cdot)$ can sharpen edges or not. Similarly, by setting $a=b>I_m$ and $a=b<I_m$, $h_T(\cdot)$ can be easily switched between the $L_2$ norm and truncated $L_2$ norm. Note that the truncated $L_2$ norm is also able to sharpen edges \cite{xu2013unnatural}. In contrast, the Welsch's penalty function does not enjoy this kind of flexibility. Different cases of $h_T(\cdot)$ are illustrated in Fig.~\ref{FigPenaltyComp}(b).

\subsection{Model}
\label{SecModel}

Given an input image $f$ and a guidance image $g$, the smoothed output image $u$ is the solution to the following objective function:
\begin{scriptsize}
\begin{equation}\label{EqObjFun}
{
  E_u(u)=\sum_{i}\left(\sum_{j\in N_d(i)}h_T(u_i-f_j) + \lambda\sum_{j\in N_s(i)}\omega_{i,j}h_T(u_i-u_j)\right),
}
\end{equation}
\end{scriptsize}
where $h_T$ is defined in Eq.(\ref{EqTruncatedHuber}); $N_d(i)$ is the $(2r_d+1)\times (2r_d+1)$ square patch centered at $i$; $N_s(i)$ is the $(2r_s+1)\times (2r_s+1)$ square patch centered at $i$; $\lambda$ is a parameter that controls the overall smoothing strength. To be clear, we adopt $\{a_d,b_d\}$ and $\{a_s,b_s\}$ to denote the parameters of $h_T(\cdot)$ in the data term and smoothness term, respectively. The guidance weight $\omega_{i,j}$ is defined as:

\begin{small}
\begin{equation}\label{EqGuidanceWeight}
{
  \omega_{i,j}=\frac{1}{(|g_i-g_j| + \delta)^\alpha},
}
\end{equation}
\end{small}


\noindent where $\alpha$ determines the sensitivity to the edges in $g$ which can be the input image, i.e., $g=f$. $|\cdot|$ represents the absolute value. $\delta$ is a small constant being set as $\delta=10^{-7}$.

The adoption of $h_T(\cdot)$ makes our model in Eq.~(\ref{EqObjFun}) to enjoy a strong flexibility. As will be shown in the following property analysis section, with different parameter settings, our model is able to achieve different smoothing behaviors, and thus it is capable of various tasks that require either edge-preserving smoothing or structure-preserving smoothing.

\subsection{Numerical Solution}

Our model in Eq.~(\ref{EqObjFun}) is not only non-convex but also non-smooth, which arises from the adopted $h_T(\cdot)$. Commonly used approaches \cite{lanckriet2009convergence,nikolova2005analysis,wang2008new,zhang2004surrogate} for solving non-convex optimization problems are not applicable. To tackle this problem, we first rewrite $h_T(\cdot)$ in a new equivalent form. By defining $\nabla^d_{i,j}=u_i-f_j$ and $\nabla^s_{i,j}=u_i-u_j$, we have:
\begin{small}
\begin{equation}\label{EqRelationWithHuberL0}
{
h_T(\nabla^\ast_{i,j})=\min_{l^\ast_{i,j}}\left\{h(\nabla^\ast_{i,j}-l^\ast_{i,j})+(b_\ast-\frac{a_\ast}{2})|l^\ast_{i,j}|_0\right\},
}
\end{equation}
\end{small}
where $\ast\in\{d,s\}$, $|l^\ast_{i,j}|_0$ is the $L_0$ norm of $l^\ast_{i,j}$. The minimum of the right side of Eq.~(\ref{EqRelationWithHuberL0}) is obtained on the condition:
\begin{small}
\begin{eqnarray}\label{EqTruncatedHuberMinCondition}
{
  l^\ast_{i,j}=\left\{\begin{array}{l}
   0, \ \ \ \ \ \ \ \ |\nabla^\ast_{i,j}|\leq b_\ast\\
   \nabla^\ast_{i,j}, \ \ |\nabla^\ast_{i,j}|>b_\ast
  \end{array}\right.
  , \ \ \ast\in\{d,s\}.
}
\end{eqnarray}
\end{small}
The detailed proof of Eq.~(\ref{EqRelationWithHuberL0}) and Eq.~(\ref{EqTruncatedHuberMinCondition}) is provided in our supplementary file. These two equations also theoretically validate our analysis in Fig.~\ref{FigPenaltyComp}(b): we have $|\nabla^\ast_{i,j}|\in[0, I_m]$ if the intensity values are in $[0, I_m]$. Then if $b>I_m$, based on Eq.~(\ref{EqRelationWithHuberL0}) and Eq.~(\ref{EqTruncatedHuberMinCondition}), we will always have $h_T(\nabla^\ast_{i,j})=h(\nabla^\ast_{i,j})$ which means $h_T(\cdot)$ degrades to $h(\cdot)$.

A new energy function is defined as:
\begin{small}
\begin{equation}\label{EqObjFunAuxUL}
{
\begin{array}{r}
   E_{ul}(u, l^d, l^s)=\sum\limits_{i,j}\left(h(\nabla^{d}_{i,j} - l^d_{i,j}) + (b_d-\frac{a_d}{2})|l^d_{i,j}|_0 \right)\\
   \ \ \ \ \ \ + \lambda\sum\limits_{i,j}\omega_{i,j}\left(h(\nabla^{s}_{i,j} - l^s_{i,j}) + (b_s-\frac{a_s}{2})|l^s_{i,j}|_0 \right)
\end{array}
}.
\end{equation}
\end{small}
Based on Eq.~(\ref{EqRelationWithHuberL0}) and Eq.~(\ref{EqTruncatedHuberMinCondition}), we then have:
\begin{small}
\begin{equation}\label{EqEnergyRelation1}
{
E_u(u)=\min_{l^\ast}E_{ul}(u, l^d, l^s),\ \ast\in\{d,s\}.
}
\end{equation}
\end{small}

Given Eq.~(\ref{EqTruncatedHuberMinCondition}) as the optimum condition of Eq.~(\ref{EqEnergyRelation1}) with respect to $l^\ast$, optimizing $E_{ul}(u, l^d, l^s)$ with respect to $u$ only involves Huber penalty function $h(\cdot)$. The problem can thus be optimized through the half-quadratic (HQ) optimization technique \cite{geman1995nonlinear,nikolova2005analysis}. More specifically, a variable $\mu^\ast (\ast\in\{d,s\})$ and a function $\psi(\mu^\ast_{i,j})$ with respect to $\mu^\ast$ exist such that:
\begin{small}
\begin{equation}\label{EqMultHQ}
{
h(\nabla^\ast_{i,j} - l^\ast_{i,j})=\min_{\mu^\ast_{i,j}}\left\{\mu^\ast_{i,j}(\nabla^\ast_{i,j}-l^\ast_{i,j})^2 + \psi(\mu^\ast_{i,j}) \right\},
}
\end{equation}
\end{small}
where the optimum is yielded on the condition:
\begin{small}
\begin{eqnarray}\label{EqMultHQCondition}
{
  \mu^\ast_{i,j}=\left\{\begin{array}{l}
   \frac{1}{2a_\ast}, \ \ \ \ \ \ \ \ \ \ \ \ \ \ \ \ \  |\nabla^\ast_{i,j} - l^\ast_{i,j}|< a_\ast\\
   \frac{1}{2|\nabla^\ast_{i,j} - l^\ast_{i,j}|}, \ \ \ \ |\nabla^\ast_{i,j} - l^\ast_{i,j}| \geq a_\ast
  \end{array}\right.
  , \ \ \ast\in\{d,s\}.
}
\end{eqnarray}
\end{small}
The detailed proof of Eq.~(\ref{EqMultHQ}) and Eq.~(\ref{EqMultHQCondition}) is provided in our supplementary file. Then we can further define a new energy function:
\begin{footnotesize}
\begin{equation}\label{EqObjFunAuxULMu}
{
\begin{array}{l}
   E_{ul\mu}(u, l^d, l^s, \mu^d, \mu^s)= \\
   \ \ \ \ \ \ \ \ \sum\limits_{i,j}\left(\mu^d_{i,j}(\nabla^{d}_{i,j} - l^d_{i,j})^2 + \psi(\mu^d_{i,j}) + (b_d-\frac{a_d}{2})|l^d_{i,j}|_0 \right) + \\
   \ \lambda\sum\limits_{i,j}\omega_{i,j}\left(\mu^s_{i,j}(\nabla^{s}_{i,j} - l^s_{i,j})^2 + \psi(\mu^s_{i,j}) + (b_s-\frac{a_s}{2})|l^s_{i,j}|_0 \right).
\end{array}
}
\end{equation}
\end{footnotesize}
 Based on Eq.~(\ref{EqMultHQ}) and Eq.~(\ref{EqMultHQCondition}), we then have:
\begin{small}
\begin{equation}\label{EqEnergyRelation2}
{
  E_{ul}(u, l^\ast)=\min\limits_{\mu^\ast}E_{ul\mu}(u, l^\ast, \mu^\ast),\ \ast\in\{d,s\}.
}
\end{equation}
\end{small}

Given Eq.~(\ref{EqMultHQCondition}) as the optimum condition of $\mu^\ast$ for Eq.~(\ref{EqEnergyRelation2}), optimizing $E_{ul\mu}(u, l^d, l^s, \mu^d, \mu^s)$ with respect to $u$ only involves the $L_2$ norm penalty function, which has a closed-form solution. However, since the optimum conditions in Eq.~(\ref{EqTruncatedHuberMinCondition}) and Eq.~(\ref{EqMultHQCondition}) both involve $u$, therefore, the final solution $u$ can only be obtained in an iterative manner. Assuming we have got $u^k$, then $(l^\ast)^{k}$ and $(\mu^\ast)^{k}, (\ast\in\{s,d\})$ can be updated through Eq.~(\ref{EqTruncatedHuberMinCondition}) and Eq.~(\ref{EqMultHQCondition}) with $u^k$. Finally, $u^{k+1}$ is obtained with:
\begin{small}
\begin{equation}\label{EqIterativeSolution}
{
   u^{k+1}=\min_{u}E_{ul\mu}\left(u, (l^\ast)^k, (\mu^\ast)^k\right),
}
\end{equation}
\end{small}
Eq.(\ref{EqIterativeSolution}) has a close-form solution as:
\begin{small}
\begin{equation}\label{EqCloseFormSolution}
{
   u^{k+1}=\left(\mathcal{A}^k - 2\lambda\mathcal{W}^k\right)^{-1}\left(D^k + 2\lambda S^k\right),
}
\end{equation}
\end{small}
where $\mathcal{W}^k$ is an affinity matrix with $\mathcal{W}^k_{i,j}=\omega_{i,j}(\mu^s_{i,j})^k$, $\mathcal{A}^k$ is a diagonal matrix with $\mathcal{A}^k_{ii}=\sum_{j\in N_d(i)}(\mu^d_{i,j})^k + 2\lambda\sum_{j\in N_s(i)}\omega_{i,j}(\mu^s_{i,j})^k$, $D^k$ is a vector with $D^k_i=\sum_{j\in N_d(i)}(\mu^d_{i,j})^k(f_j+(l^d_{i,j})^k)$ and $S^k$ is also a vector with $S^k_i=\sum_{j\in N_s(i)}\omega_{i,j}(\mu^s_{i,j})^k(l^s_{i,j})^k$.

The above optimization procedure monotonically decreases the value of $E_u(u)$ in each step, its convergence is theoretically guaranteed. Given $u^k$ in the $k$th iteration and $\ast\in\{s,d\}$, then for any $u$, we have:
\begin{small}
\begin{equation}\label{EqEnergyRelationTruncation}
{
  E_u(u)\leq E_{ul}(u, (l^\ast)^k),\ E_u(u^k)=E_{ul}(u^k, (l^\ast)^k),
}
\end{equation}
\begin{eqnarray}\label{EqEnergyRelationHQ}
{\left\{
\begin{array}{l}
  E_{ul}(u,(l^\ast)^k)\leq E_{ul\mu}(u, (l^\ast)^k, (\mu^\ast)^k)\\
  E_{ul}(u^k,(l^\ast)^k)=E_{ul\mu}(u^k, (l^\ast)^k, (\mu^\ast)^k)
\end{array}\right. .
}
\end{eqnarray}
\end{small}

Given $(l^\ast)^k$ has been updated through Eq.~(\ref{EqTruncatedHuberMinCondition}), Eq.~(\ref{EqEnergyRelationTruncation}) is based on Eq.~(\ref{EqEnergyRelation1}) and Eq.~(\ref{EqRelationWithHuberL0}). After $(\mu^\ast)^k$ has been updated through Eq.~(\ref{EqMultHQCondition}), Eq.~(\ref{EqEnergyRelationHQ}) is based on Eq.~(\ref{EqEnergyRelation2}) and Eq.~(\ref{EqMultHQ}). We now have:
\begin{small}
\begin{equation}\label{EqEnergyDecrease1}
{
\begin{array}{l}
  E_{ul}(u^{k+1},(l^\ast)^k)\leq E_{ul\mu}(u^{k+1}, (l^\ast)^k, (\mu^\ast)^k)\\
  \leq E_{ul\mu}(u^{k}, (l^\ast)^k, (\mu^\ast)^k)=E_{ul}(u^{k},(l^\ast)^k),
\end{array}
}
\end{equation}
\end{small}
the first and second inequalities follow from Eq.~(\ref{EqEnergyRelationHQ}) and Eq.~(\ref{EqIterativeSolution}), respectively. We finally have:
\begin{small}
\begin{equation}\label{EqEnergyDecrease2}
{
  E_u(u^{k+1})\leq E_{ul}(u^{k+1}, (l^\ast)^k)\leq E_{ul}(u^{k}, (l^\ast)^k)=E_{u}(u^k),
}
\end{equation}
\end{small}
the first and second inequalities follow from Eq.~(\ref{EqEnergyRelationTruncation}) and Eq.~(\ref{EqEnergyDecrease1}), respectively. Since the value of $E_u(u)$ is bounded from below, Eq.~(\ref{EqEnergyDecrease2}) indicates that the convergence of our iterative scheme is theoretically guaranteed.

The above optimization procedure is iteratively performed $N$ times to get the output $u^N$.  In all our experiments, we set $u^0=f$, which is able to produce promising results in each application. Our optimization procedure is summarized in Algorithm~\ref{Alg}.

\begin{algorithm}[t]
\caption {Image Smoothing via Non-convex Non-smooth Optimization}\label{Alg}
\begin{algorithmic}[1]
\REQUIRE
Input image $f$, guide image $g$, iteration number $N$, parameter $\lambda, \alpha, a_\ast, b_\ast, r_\ast$, $u^0\leftarrow f$, with $\ast\in\{d,s\}$\\

\FOR{$k=0:N$}
\STATE With $u^k$, compute $(\nabla^\ast_{i,j})^k$, update $(l^\ast_{i,j})^k$ according to Eq.~(\ref{EqTruncatedHuberMinCondition})
\STATE With $(l^\ast_{i,j})^k$, update $(\mu^\ast_{i,j})^k$ according to Eq.~(\ref{EqMultHQCondition})
\STATE With $(l^\ast_{i,j})^k$ and $(\mu^\ast_{i,j})^k$, solve for $u^{k+1}$ according to Eq.~(\ref{EqIterativeSolution}) (or Eq.~(\ref{EqCloseFormSolution}))
\ENDFOR
\ENSURE
Smoothed image $u^{N+1}$
\end{algorithmic}
\end{algorithm}

\subsection{Property Analysis}
\label{SecPropertyAnalysis}

With different parameter settings, the strong flexibility of $h_T(\cdot)$ makes our model able to achieve various smoothing behaviors. First, we show that some classical approaches can be viewed as special cases of our model. For example, by setting $a_d=b_d>I_m, a_s=\epsilon,b_s>I_m,\alpha=0,r_d=0,r_s=1$, our model is an approximation of the TV model \cite{rudin1992nonlinear} which is a representative edge-preserving smoothing operator. If we set $\alpha=0.2,g=f$ with other parameters the same as above, then the first iteration of Algorithm~\ref{Alg} will be the WLS smoothing \cite{farbman2008edge} which performs well in handling gradient reversals and halos in image detail enhancement and HDR tone mapping. With parameters $a_d=\epsilon,b_d>I_m,a_s=\epsilon,b_s>I_m,\alpha=0,r_d=0,r_s=1$, our model can yield very close smoothing natures as the TV-$L_1$ model \cite{buades2010fast} which is classical for structure-preserving smoothing.

\begin{figure*}
\centering
\subfigure[]
{
\includegraphics[width=0.13\linewidth]{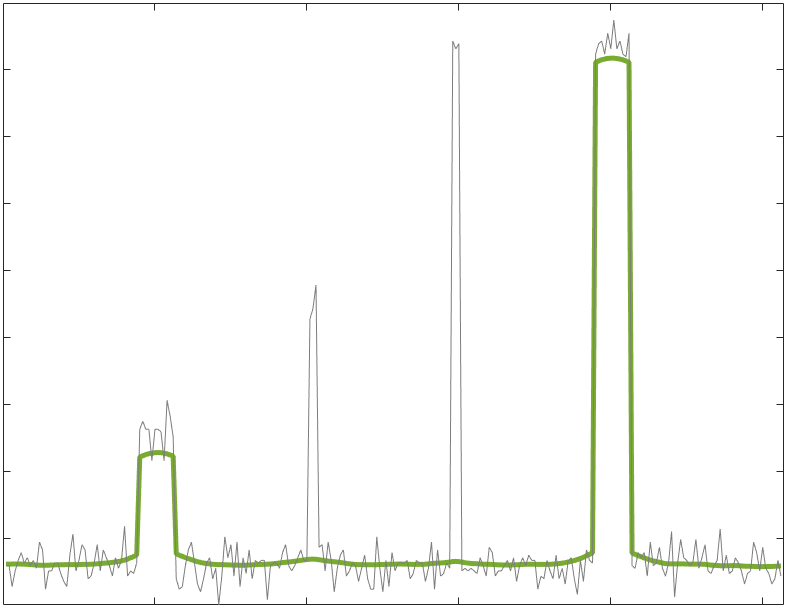}
}
\subfigure[]
{
\includegraphics[width=0.13\linewidth]{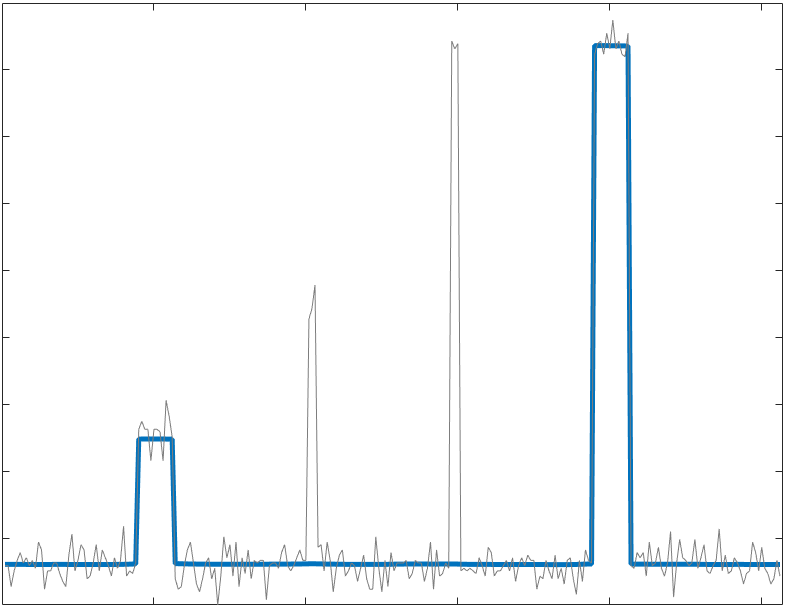}
}
\subfigure[]
{
\includegraphics[width=0.13\linewidth]{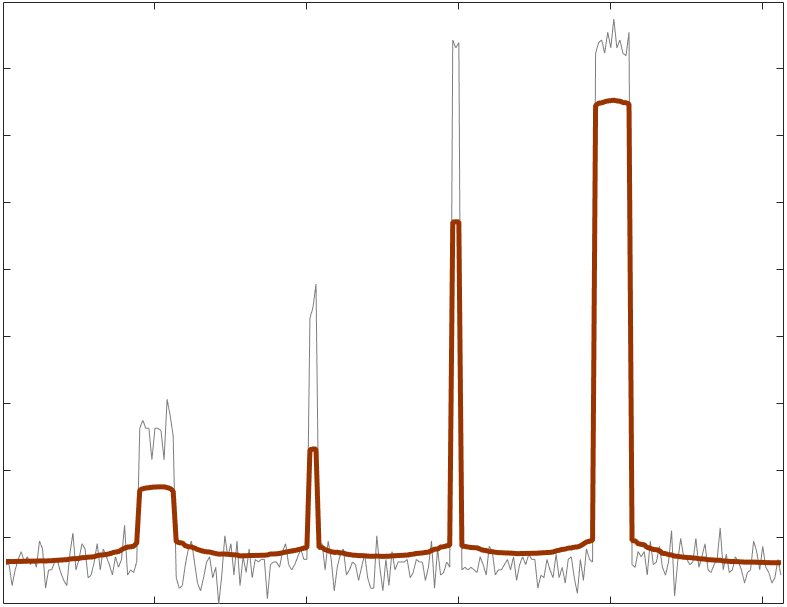}
}
\subfigure[]
{
\includegraphics[width=0.13\linewidth]{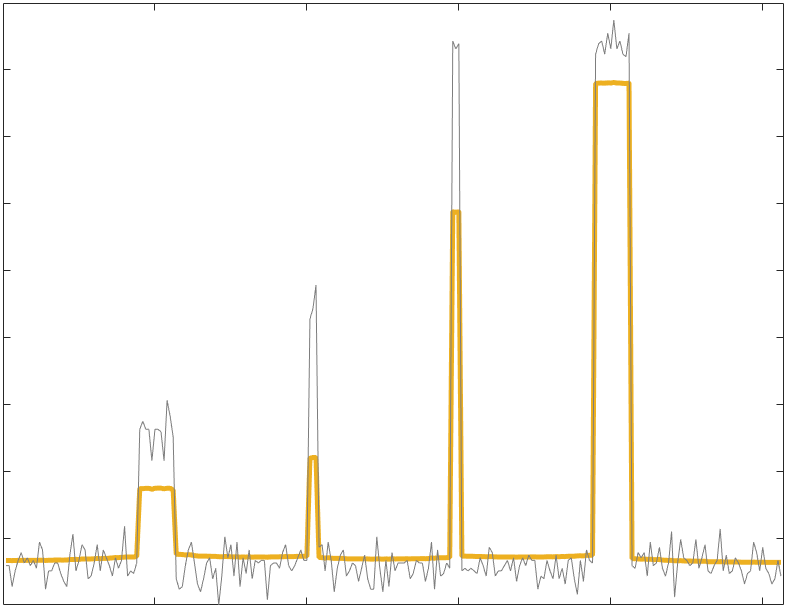}
}
\subfigure[]
{
\includegraphics[width=0.13\linewidth]{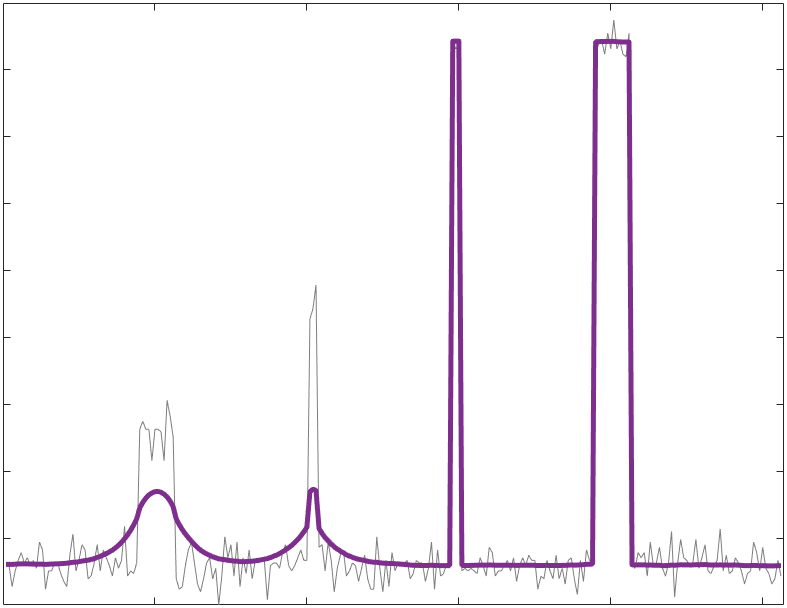}
}
\subfigure[]
{
\includegraphics[width=0.13\linewidth]{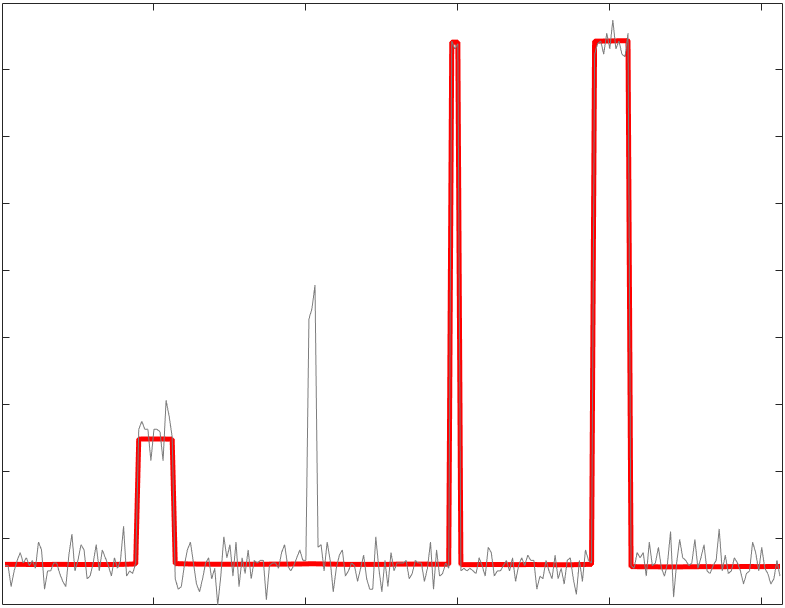}
}
\caption{1D signal with structures of different scales and amplitudes. Smoothing result of (a) TV-$L_1$ smoothing \cite{buades2010fast}, (c) WLS \cite{farbman2008edge}, (e) SD filter \cite{ham2015robust}, our results in (b), (d) and (f).}\label{Fig1DComp}
\end{figure*}
\begin{figure*}[!ht]
  \centering
  \subfigure[]
  {
  \includegraphics[width=0.22\linewidth]{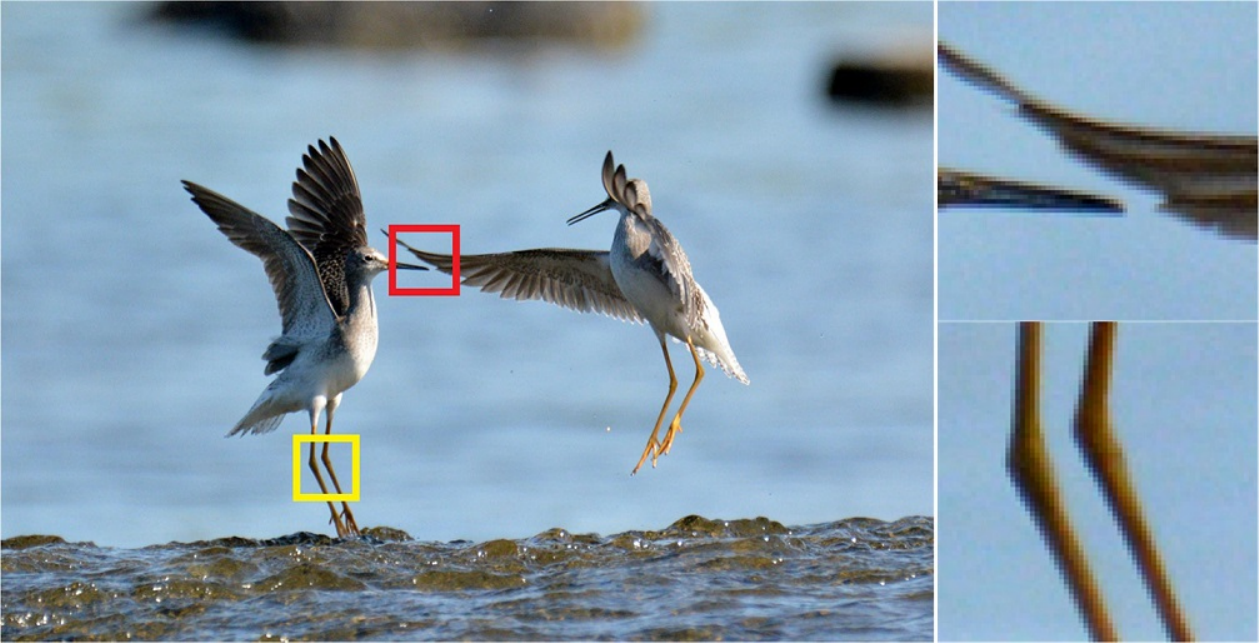}
  }
  \subfigure[]
  {
  \includegraphics[width=0.2775\linewidth]{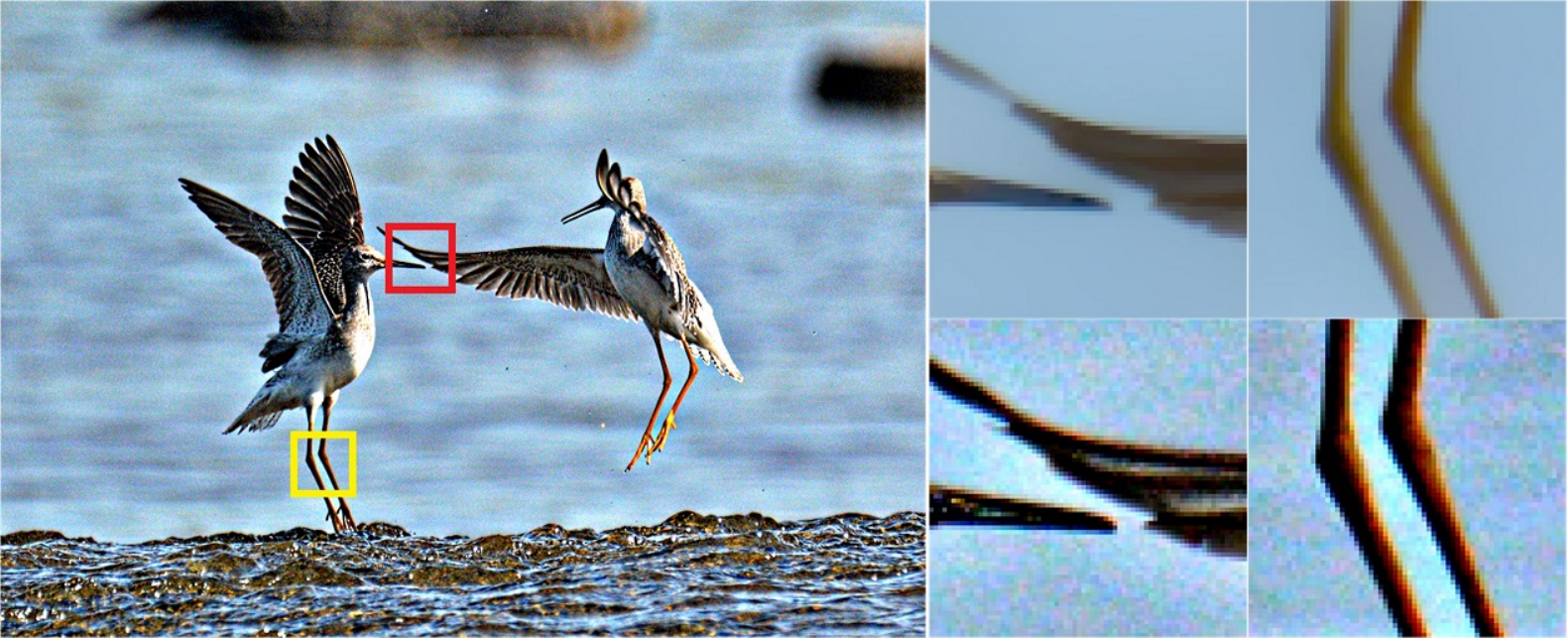}
  }
  \subfigure[]
  {
  \includegraphics[width=0.2775\linewidth]{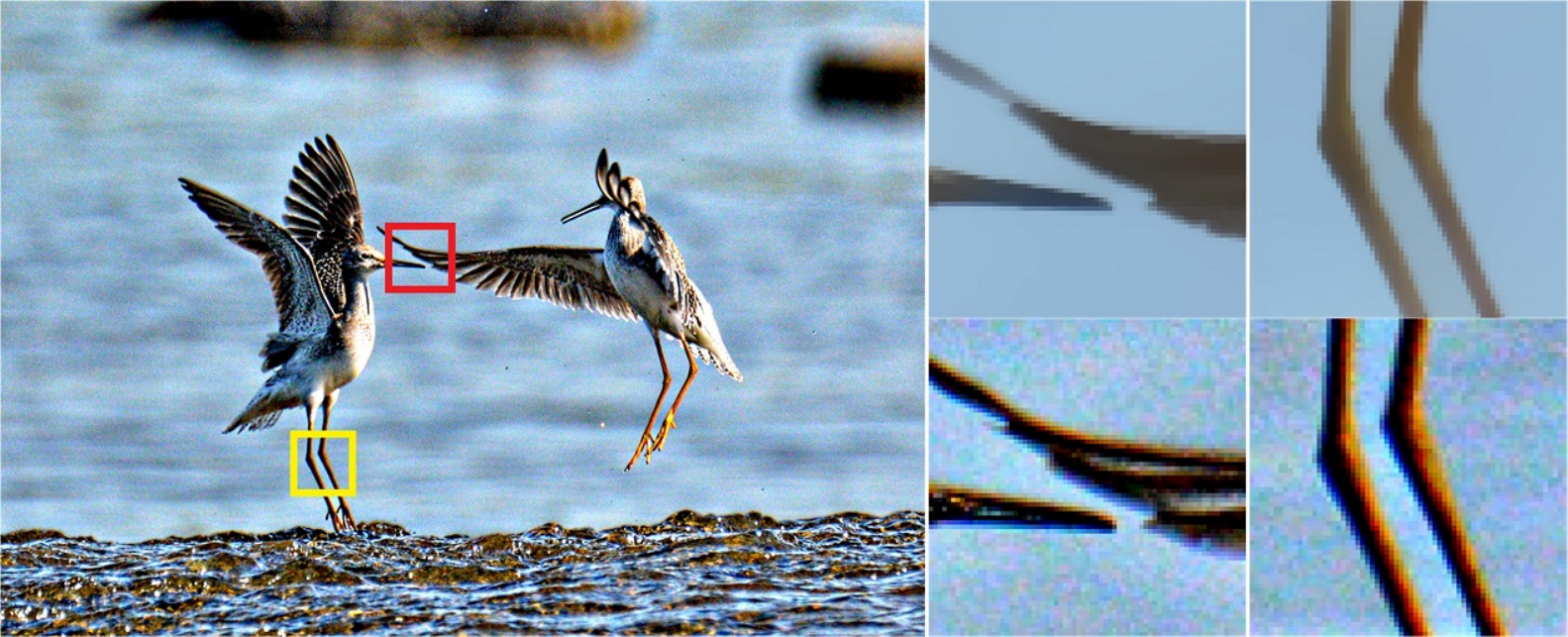}
  }
  \caption{Image detail enhancement results of different approaches. (a) Input image. Result of (b) WLS \cite{farbman2008edge} and (c) our method. The upper parts of each close-up in (b) and (c) correspond to the patches in the smoothed image.}\label{FigMyVsWLS}
\end{figure*}
\begin{figure*}[!ht]
  \centering
  \subfigure[]
  {
  \includegraphics[width=0.2\linewidth]{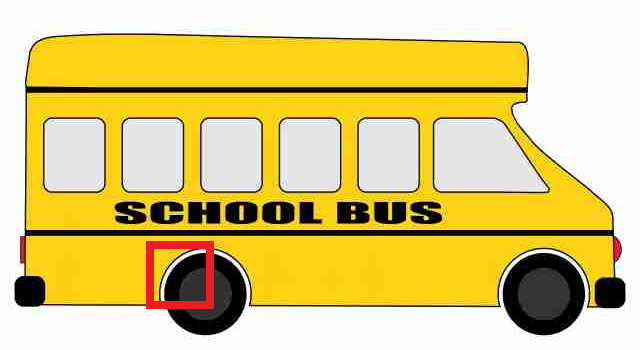}
  }
  \subfigure[]
  {
  \includegraphics[width=0.2\linewidth]{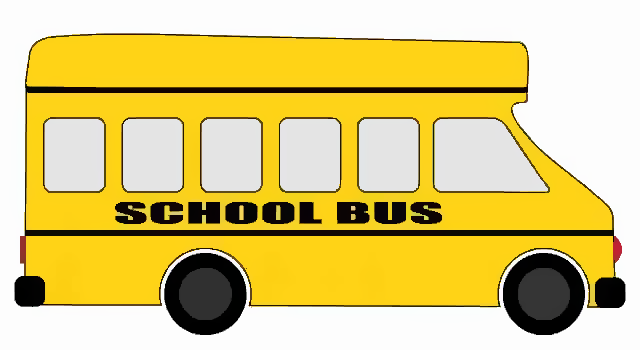}
  }
  \subfigure[]
  {
  \includegraphics[width=0.1\linewidth]{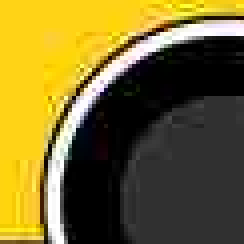}
  }
  \subfigure[]
  {
  \includegraphics[width=0.1\linewidth]{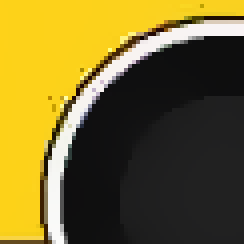}
  }
  \subfigure[]
  {
  \includegraphics[width=0.1\linewidth]{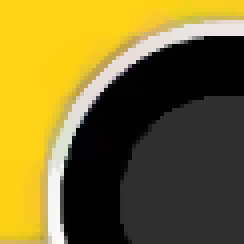}
  }
  \subfigure[]
  {
  \includegraphics[width=0.1\linewidth]{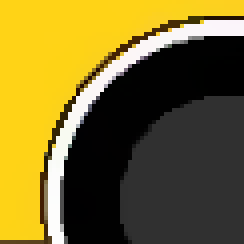}
  }
  \caption{Clip-art compression artifacts removal results of different approaches. (a) Input image. (b) Our result. Close-ups of (c) input image and results of (d) SD filter \cite{ham2015robust}, (e) our method with the structure-preserving parameter setting, (f) our method with the edge-preserving and structure-preserving parameter setting.}\label{FigMyVsSDFilter}
\end{figure*}
\begin{figure}
  \centering
  \subfigure[]
  {
  \includegraphics[width=0.27\linewidth]{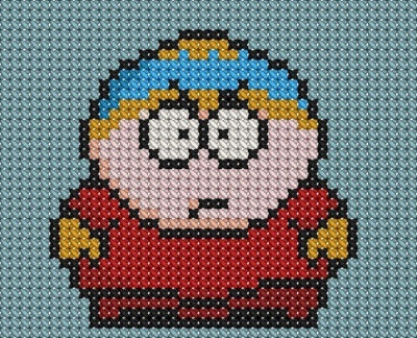}
  }
  \subfigure[]
  {
  \includegraphics[width=0.27\linewidth]{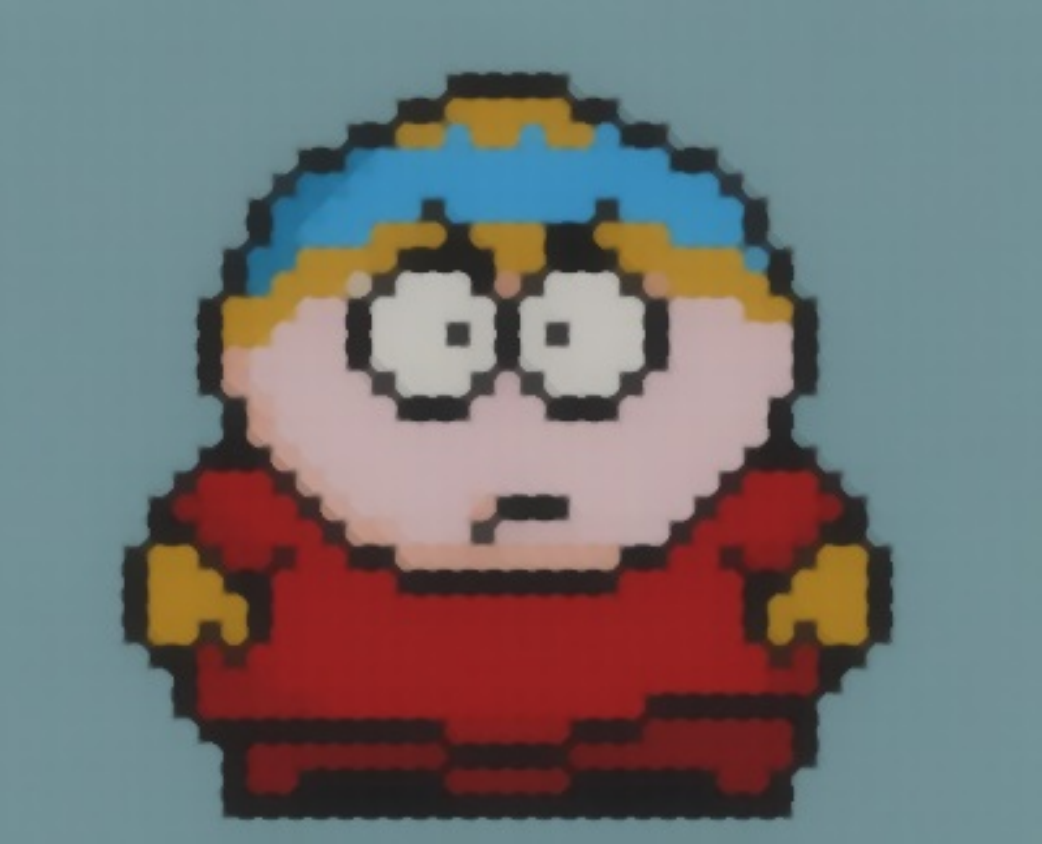}
  }
  \subfigure[]
  {
  \includegraphics[width=0.27\linewidth]{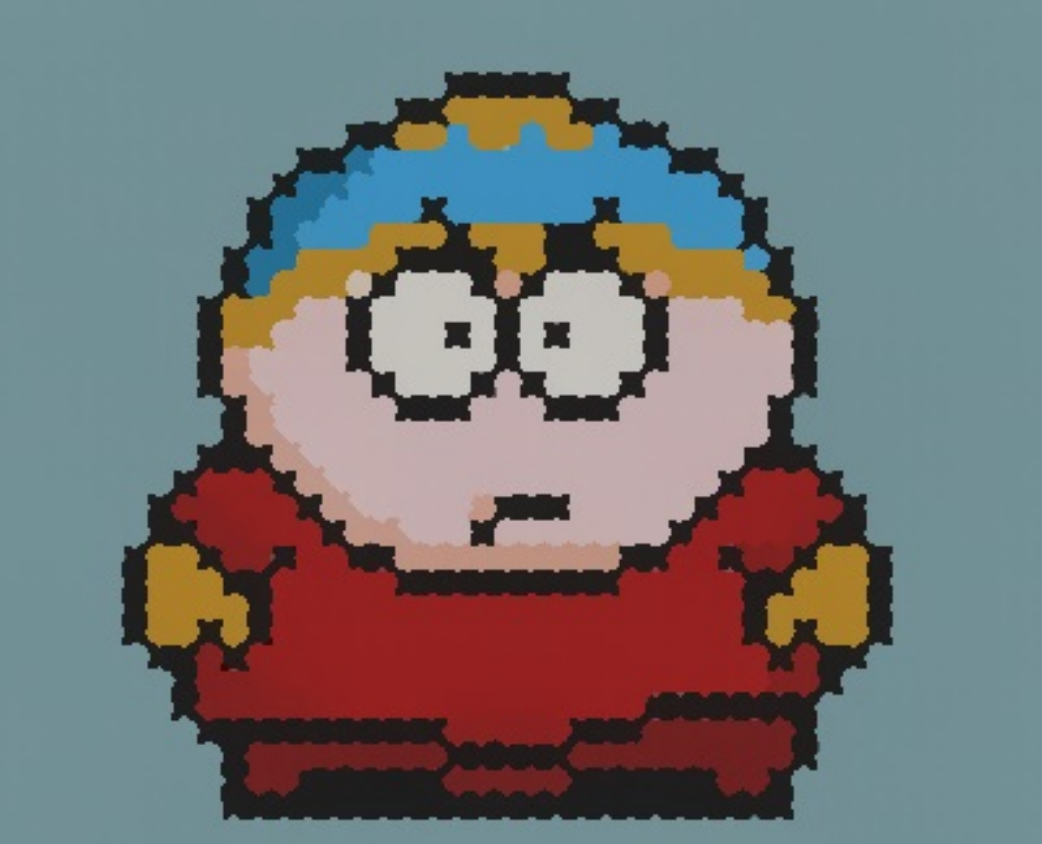}
  }
  \caption{Texture smoothing results of different approaches. (a) Input image. Result of (b) TV-$L_1$ smoothing \cite{buades2010fast}, and (e) our method.}\label{FigMyVsTVL1}
\end{figure}

For different kinds of applications, our model can produce better results than the special cases mentioned above. To be convenient, we first start with the tasks in the fourth group which require structure-preserving smoothing. For these tasks, the parameters are set as $a_d=\epsilon,b_d>I_m,a_s=\epsilon,b_s>I_m,r_d=r_s,\alpha=0.5,g=f$. This parameter setting has the following two advantages: first, the setting $a_d=\epsilon,b_d>I_m,a_s=\epsilon,b_s>I_m$ enables our model to have the structure-preserving property similar to that of the TV-$L_1$ model; second, the guidance weight with $\alpha=0.5,g=f$ can make our model to obtain sharper edges in the results than the TV-$L_1$ model does. We illustrate this with 1D smoothing results in Fig.~\ref{Fig1DComp}(a) and (b). Fig.~\ref{FigMyVsTVL1}(b) and (c) further show a comparison of image texture removal results. As shown in the figure, both the TV-$L_1$ model and our model can properly remove the small textures, however, edges in our result are much sharper than that in the result of the TV-$L_1$ model. The typical values for $r_d=r_s$ are $1\sim3$ depending on the texture size. $\lambda$ is usually smaller than 1. Larger $r_d,r_s,\lambda$ can lead larger structures to be removed. The iteration number is set as $N=10$.

When dealing with image detail enhancement and HDR tone mapping in the first group, one way is to set the parameters so that our model can perform WLS smoothing. In contrast, we can further make use of the structure-preserving property of our model to produce better results. The parameters are set as follows: $a_d=\epsilon,b_d>I_m,a_s=\epsilon,b_s>I_m,r_d=r_s,\alpha=0.2,g=f$. This kind of parameter setting is based on the following observation in our experiments: when we adopt $N=1$ and set $\lambda$ to a large value, the amplitudes of different structures will decrease at different rates, i.e., the amplitudes of small structures can have a larger decrease than the large ones, as illustrated in Fig.~\ref{Fig1DComp}(d). At the same time, edges are neither blurred nor sharpened. This kind of smoothing behavior is desirable for image detail enhancement and HDR tone mapping. As a comparison, Fig.~\ref{Fig1DComp}(c) shows the smoothing result of the WLS smoothing. As can be observed from the figures, our method can better preserve the edges (see the bottom of the 1D signals in Fig.~\ref{Fig1DComp}(c) and (d)). Fig.~\ref{FigMyVsWLS}(b) and (c) further show a comparison of image detail enhancement results. We fix $r_d=r_s=2$ and vary $\lambda$ to control the smoothing strength. $\lambda$ for the tasks in the first group is usually much larger than that for the ones in the fourth group, for example, the result in Fig.~\ref{FigMyVsWLS}(c) is generated with $\lambda=20$.

\begin{figure*}[!ht]
  \centering
  \subfigure[]
  {
  \includegraphics[width=0.135\linewidth]{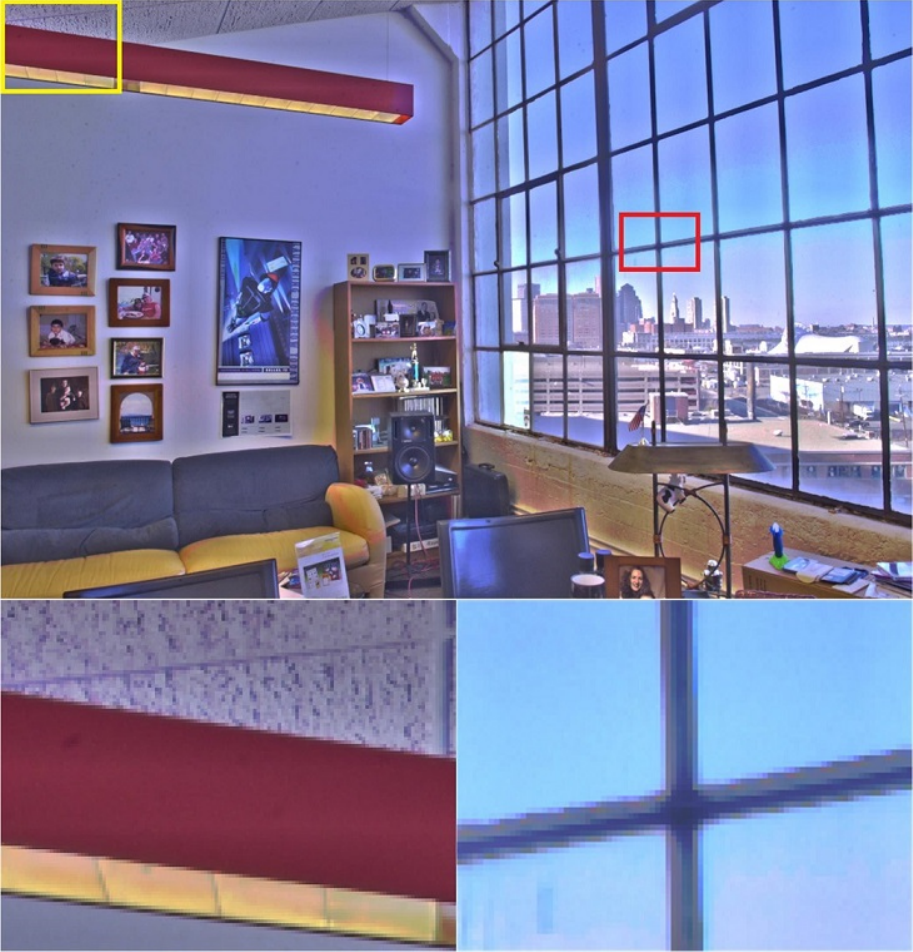}
  }
  \subfigure[]
  {
  \includegraphics[width=0.135\linewidth]{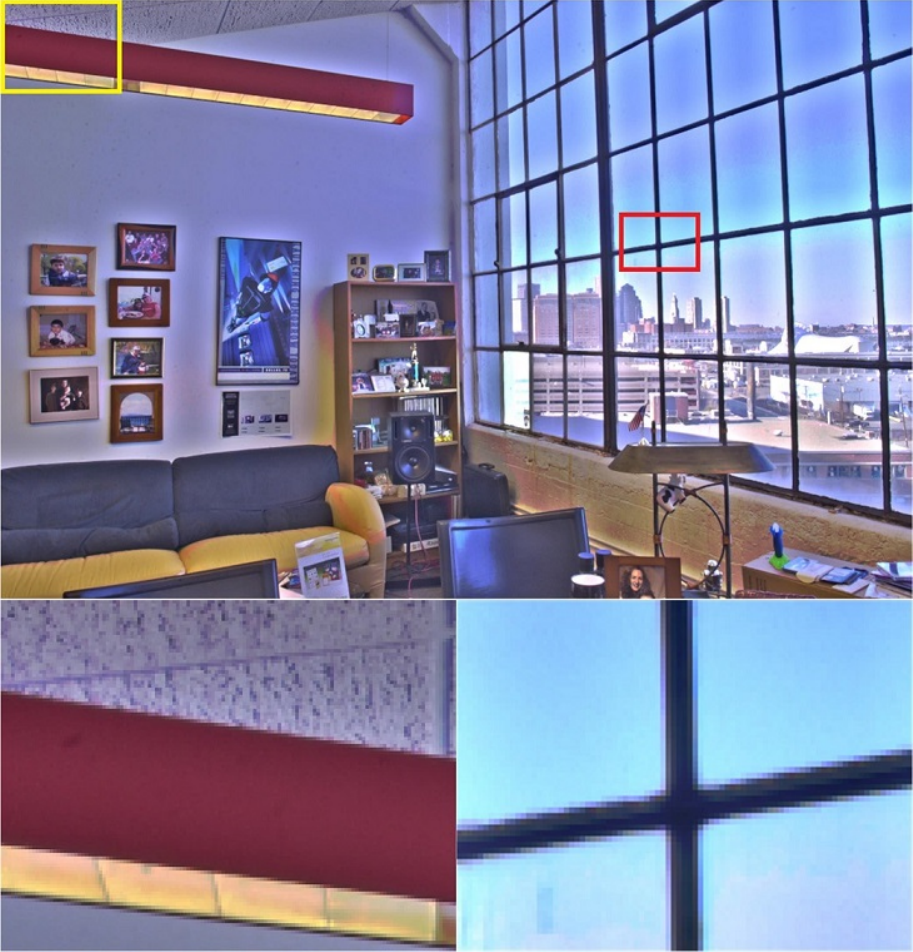}
  }
  \subfigure[]
  {
  \includegraphics[width=0.135\linewidth]{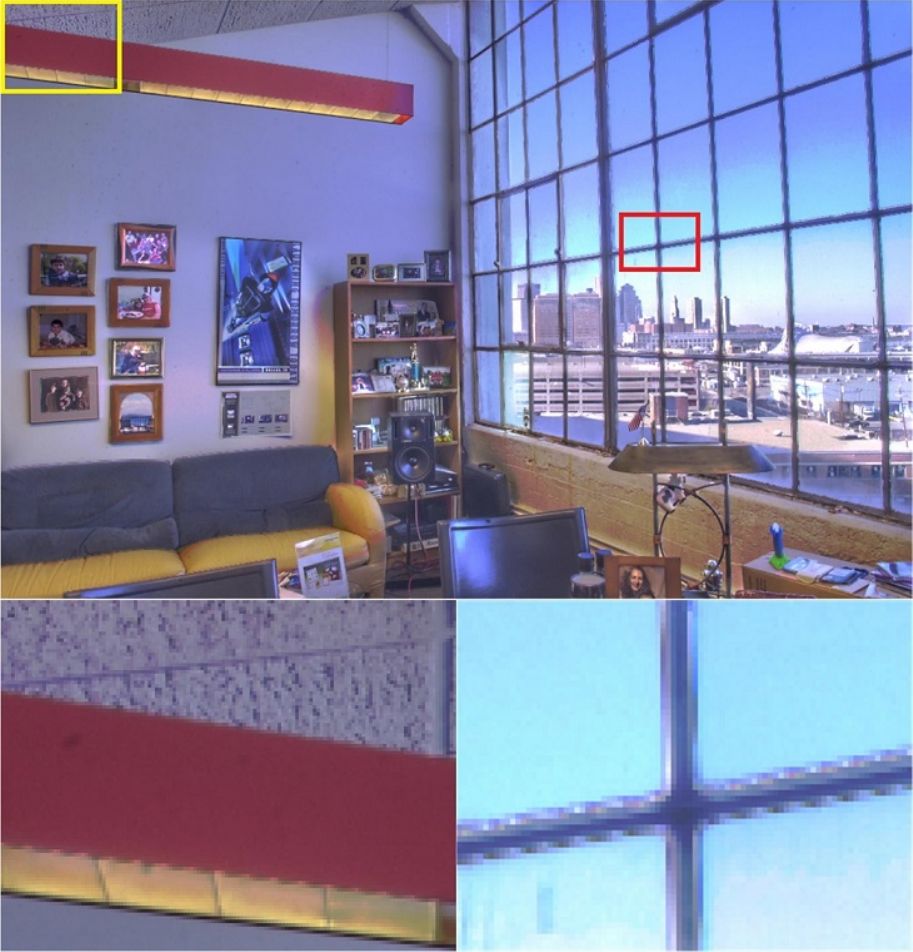}
  }
  \subfigure[]
  {
  \includegraphics[width=0.135\linewidth]{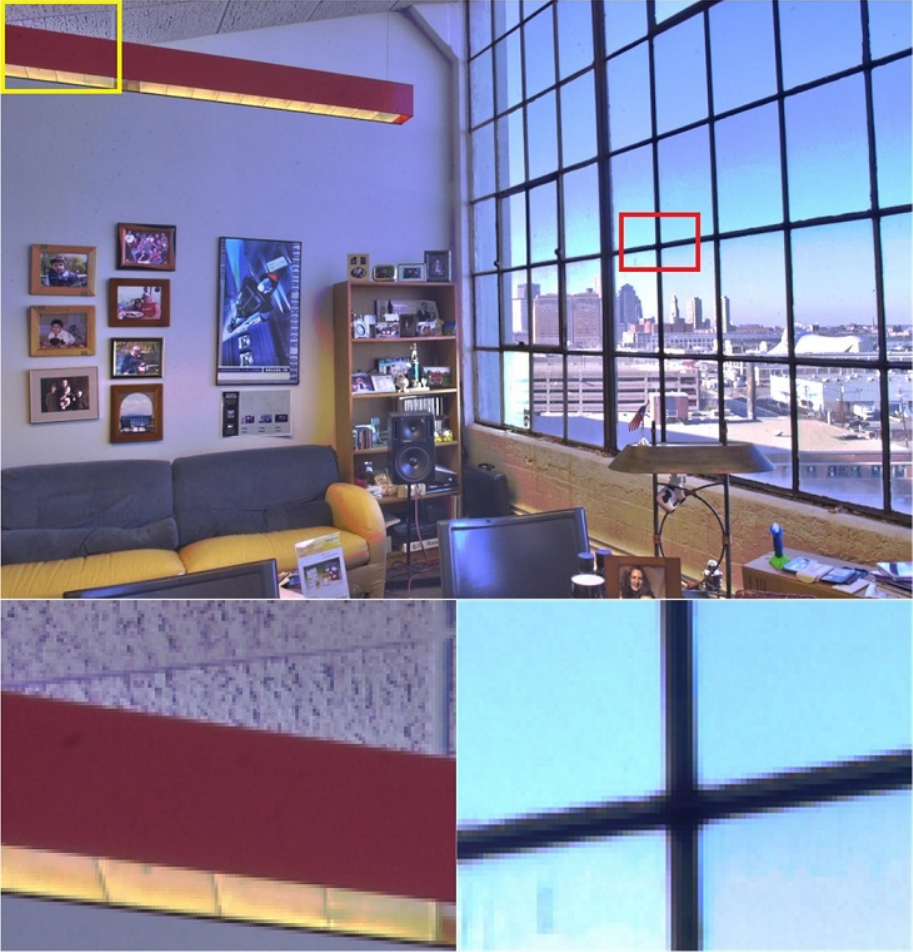}
  }
  \subfigure[]
  {
  \includegraphics[width=0.135\linewidth]{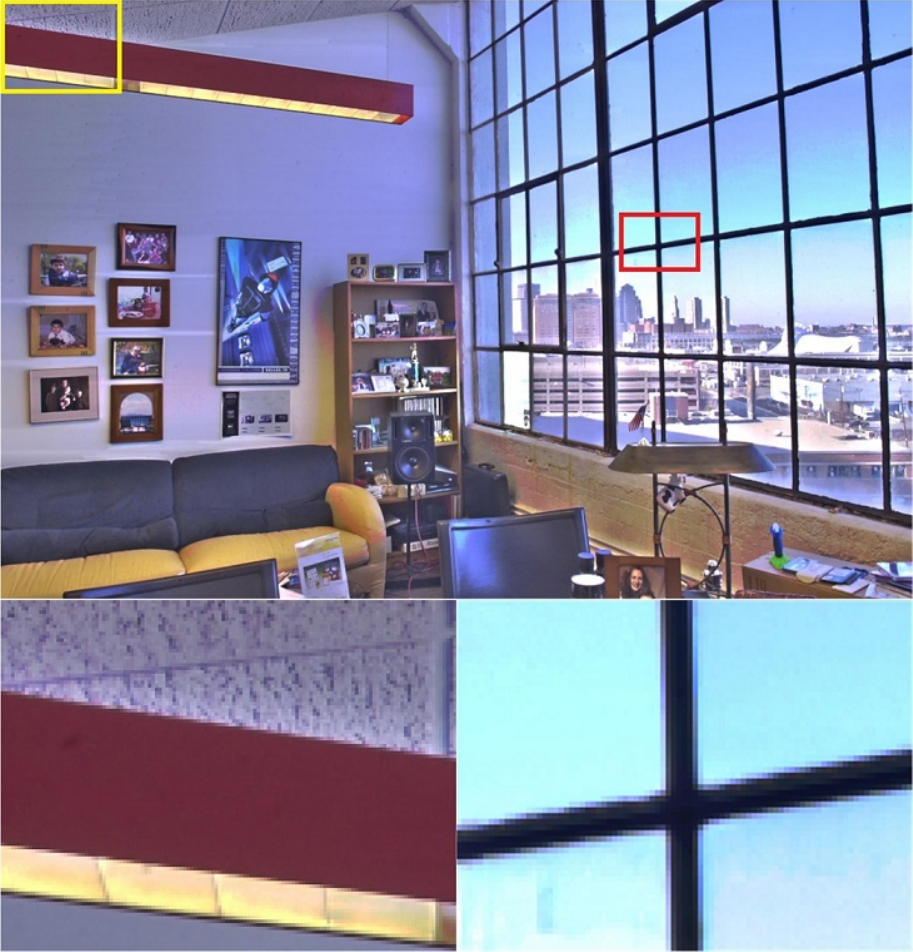}
  }
  \subfigure[]
  {
  \includegraphics[width=0.135\linewidth]{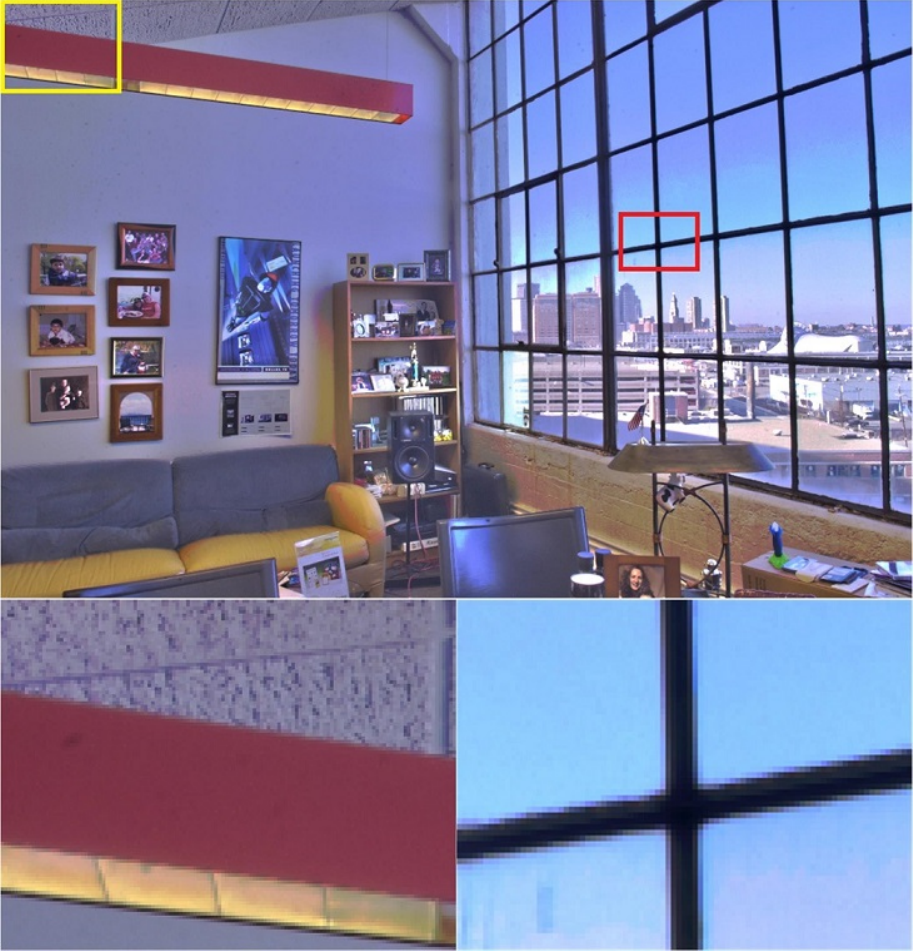}
  }
  \caption{HDR tone mapping results of different approaches. Result of (a) BF \cite{tomasi1998bilateral}, (b) GF \cite{he2013guided}, (c) $L_0$ norm smoothing \cite{xu2011image}, (d) WLS \cite{farbman2008edge}, (e) SG-WLS \cite{liu2017semi} and (f) our method.}\label{FigHDRToneMapping}
\end{figure*}

\begin{figure*}
  \centering
  \subfigure[]
  {
  \includegraphics[width=0.135\linewidth]{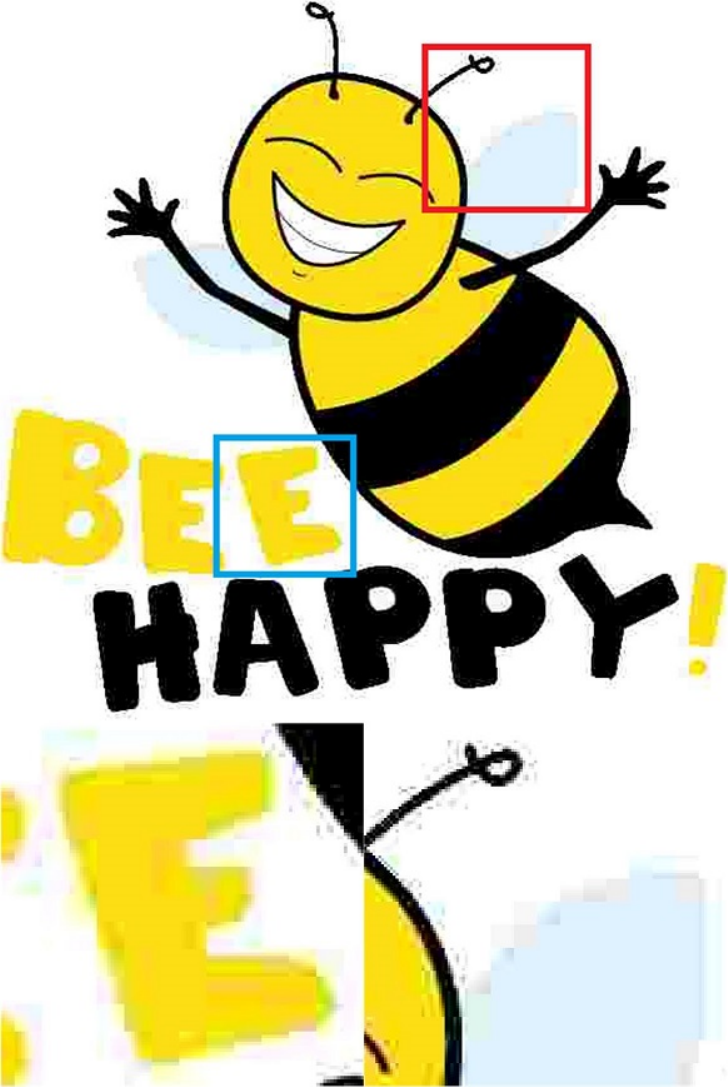}
  }
  \subfigure[]
  {
  \includegraphics[width=0.135\linewidth]{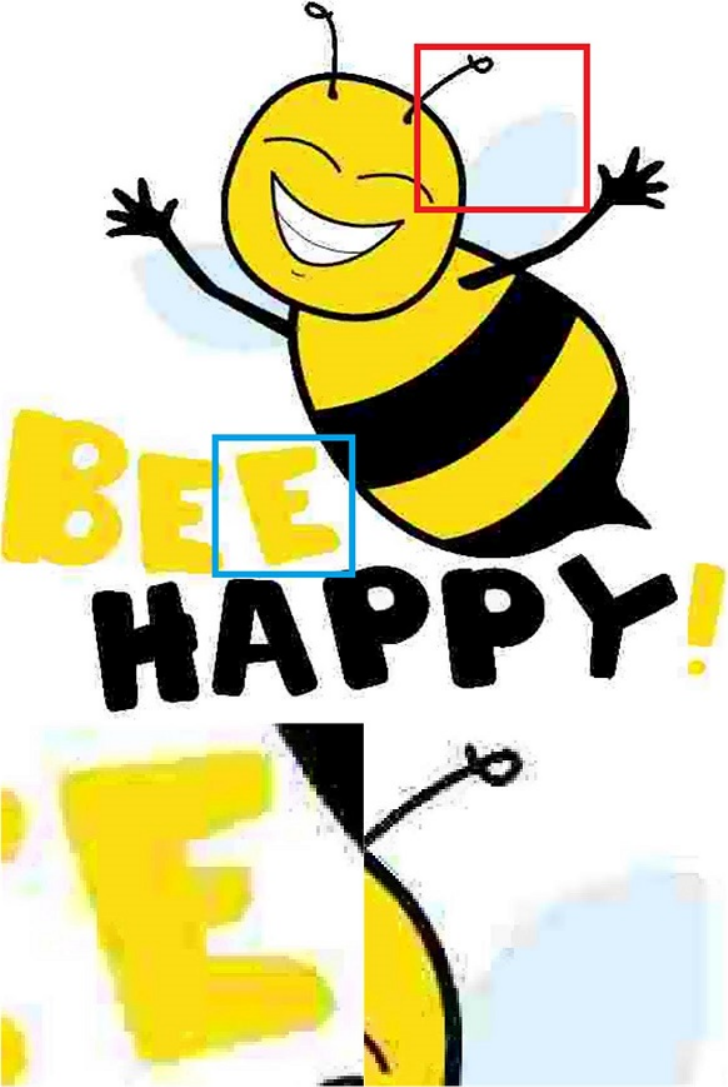}
  }
  \subfigure[]
  {
  \includegraphics[width=0.135\linewidth]{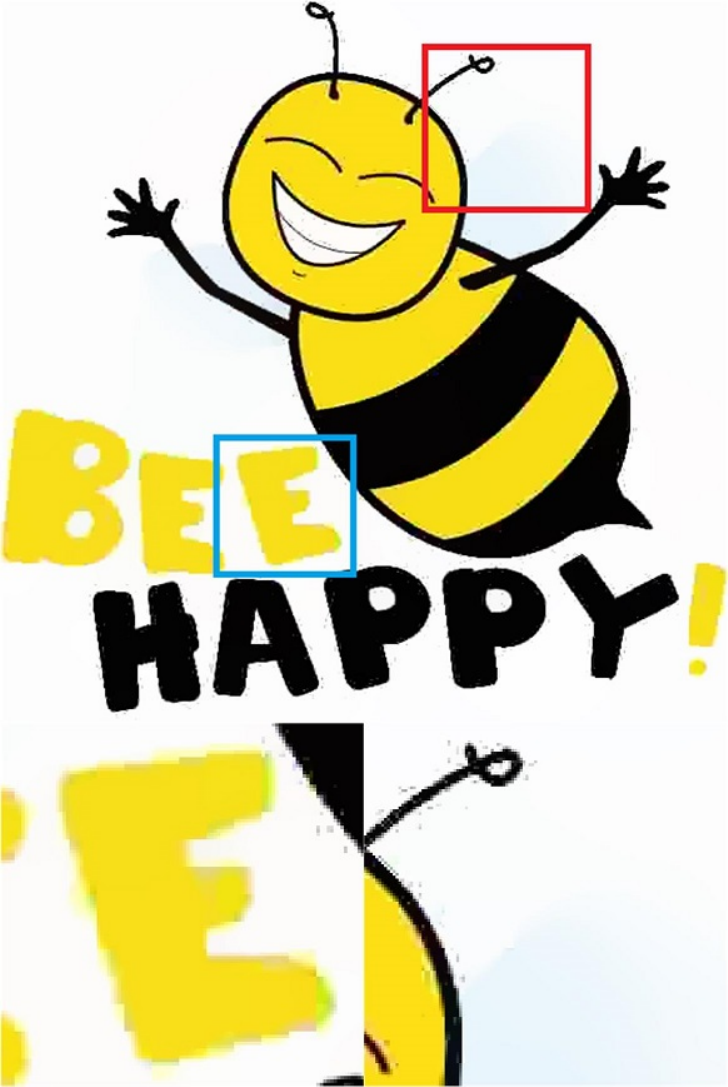}
  }
  \subfigure[]
  {
  \includegraphics[width=0.135\linewidth]{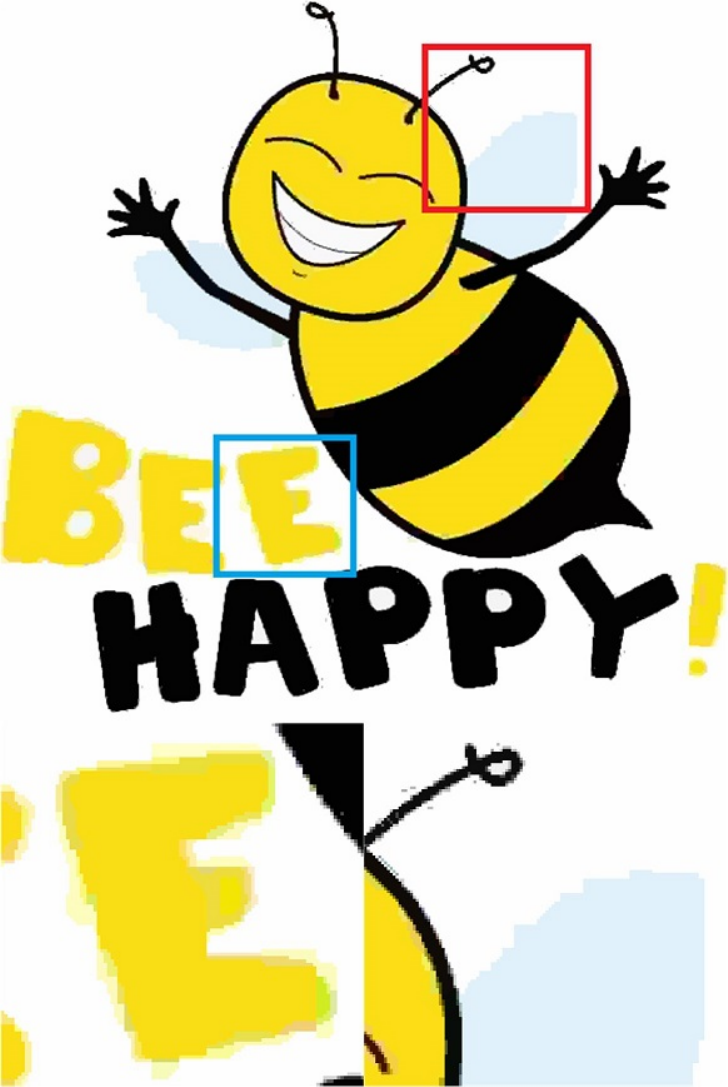}
  }
  \subfigure[]
  {
  \includegraphics[width=0.135\linewidth]{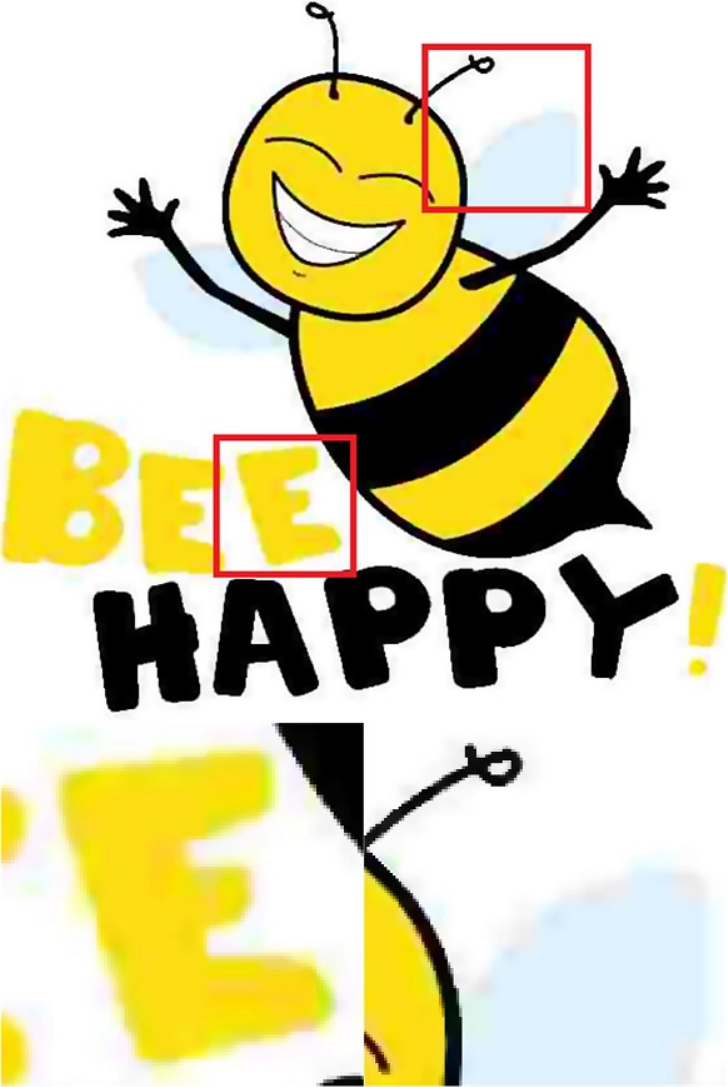}
  }
  \subfigure[]
  {
  \includegraphics[width=0.135\linewidth]{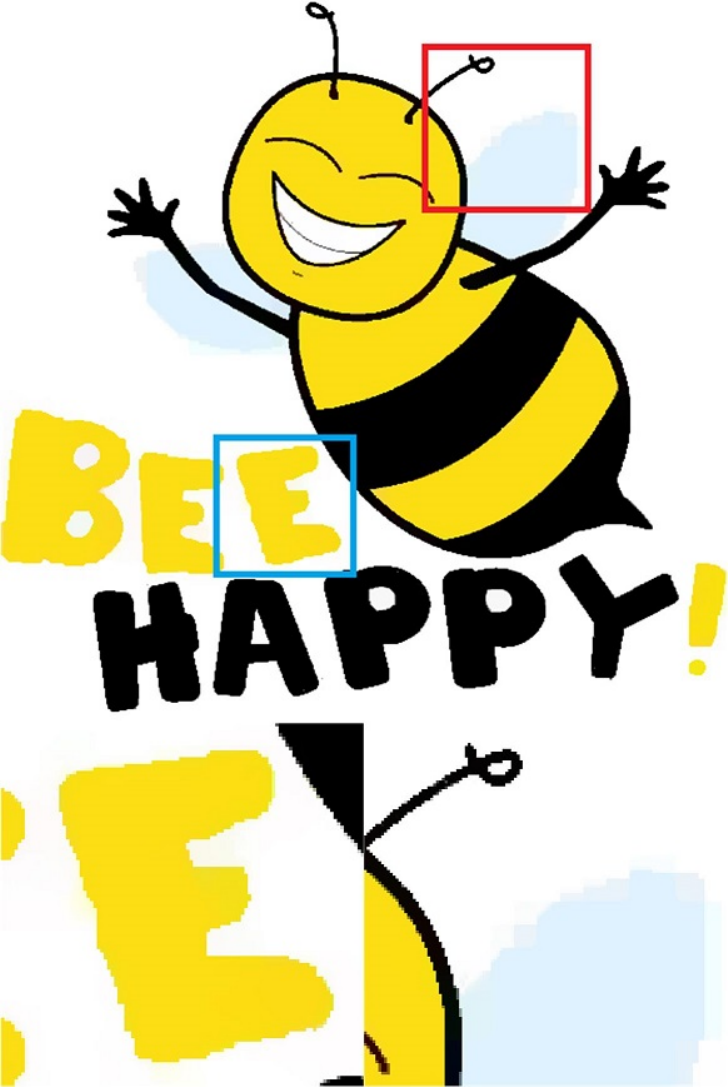}
  }
  \caption{Clip-art compression artifacts removal results of different methods. (a) Input compressed image. Result of (b) the approach proposed by Wang et~al. \cite{wang2006deringing}, (c) $L_0$ norm smoothing \cite{xu2011image}, (d) region fusion approach \cite{nguyen2015fast}, (e) BTF \cite{cho2014bilateral} and (f) our method.}\label{FigClipArt}
\end{figure*}

To sharpen edges that is required by the tasks in the second and the third groups, we can set $b_s<I_m$ in the smoothness term. In addition, we further set other parameters as $a_d=\epsilon, b_d<I_m, a_s=\epsilon$. The truncation $b_d<I_m$ in the data term can help our model to be robust against the outliers in the input image, for example, the noise in the no flash image and low-quality depth map. The truncation $b_s<I_m$ in the smoothness term can enable our model to be an edge-preserving one. By setting $a_d=a_s=\epsilon$, our model can further enjoy the structure-preserving property. With both edge-preserving and structure-preserving smoothing natures, our model has the ability to preserve large structures with weak edges and small structures with strong edges at the same time, which is challenging but is of practical importance. Fig.~\ref{FigMyVsSDFilter}(a) illustrates this kind of case with an example of clip-art compression artifacts removal: both the thin black circle around the ``wheel'' and the gray part in the center of the ``wheel'' should be preserved. The challenge lies on two facts. On one hand, if we perform edge-preserving smoothing, the gray part will be removed because the corresponding edge is weak. Fig.~\ref{FigMyVsSDFilter}(d) shows the result of the SD filter \cite{ham2015robust}. The SD filter can properly preserve the thin black circle and sharpen the edges thanks to the adopted Welsch's penalty function, however, it fails to preserve the weak edge between the black part and the gray part. On the other hand, if we adopt structure-preserving smoothing, then the thin black circle will be smoothed due to its small structure size. Fig.~\ref{FigMyVsSDFilter}(e) shows the corresponding result of our method with the structure-preserving parameter setting described above. In contrast, our method with the edge-preserving and structure-preserving parameter setting can preserve both these two parts and sharpen the edges, as shown in Fig.~\ref{FigMyVsSDFilter}(f). Fig.~\ref{Fig1DComp}(e) and (f) also show a comparison of the SD filter and our method with 1D smoothing results. We fix $\alpha=0.5, r_d=r_s, N=10$ for the tasks in both the second and the third groups. We empirically set $b_d=b_s=0.05I_m\sim0.2I_m$ and $r_d=r_s=1\sim5$ depending on the applied task and the input noise level.

\begin{figure*}[!ht]
  \centering
  \subfigure[]
  {
  \includegraphics[width=0.122\linewidth]{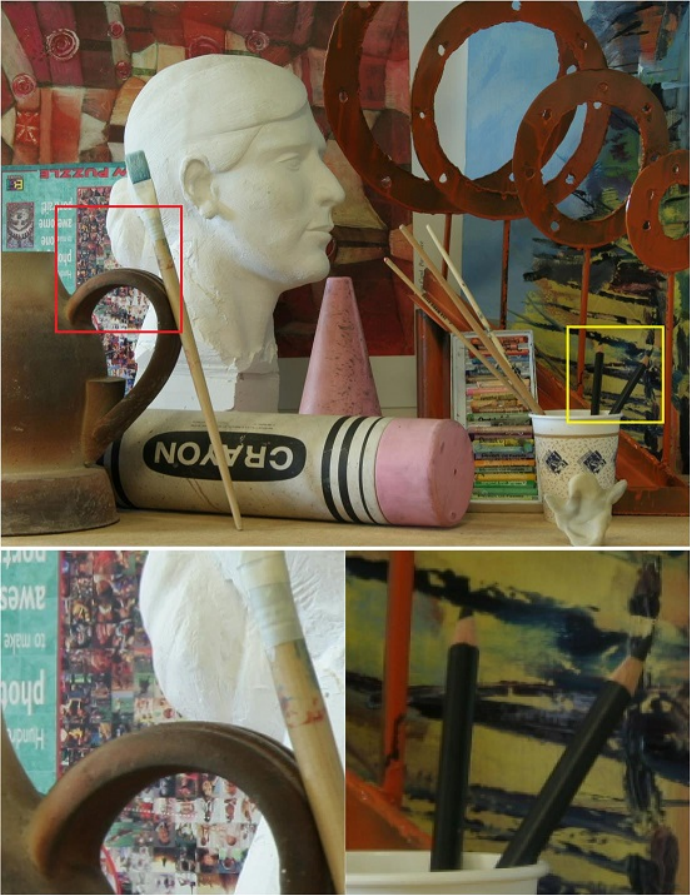}
  }
  \subfigure[]
  {
  \includegraphics[width=0.122\linewidth]{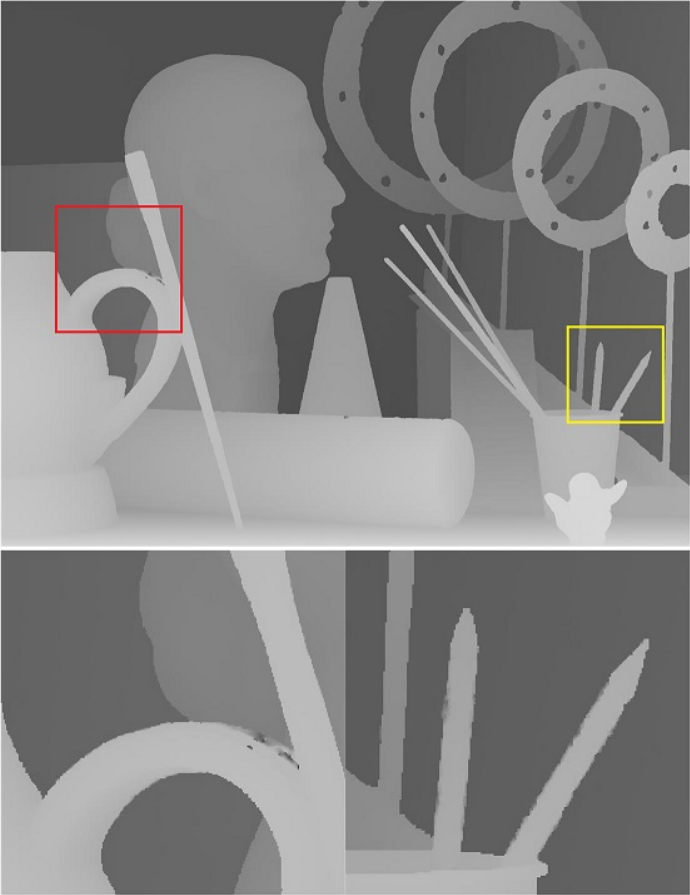}
  }
  \subfigure[]
  {
  \includegraphics[width=0.122\linewidth]{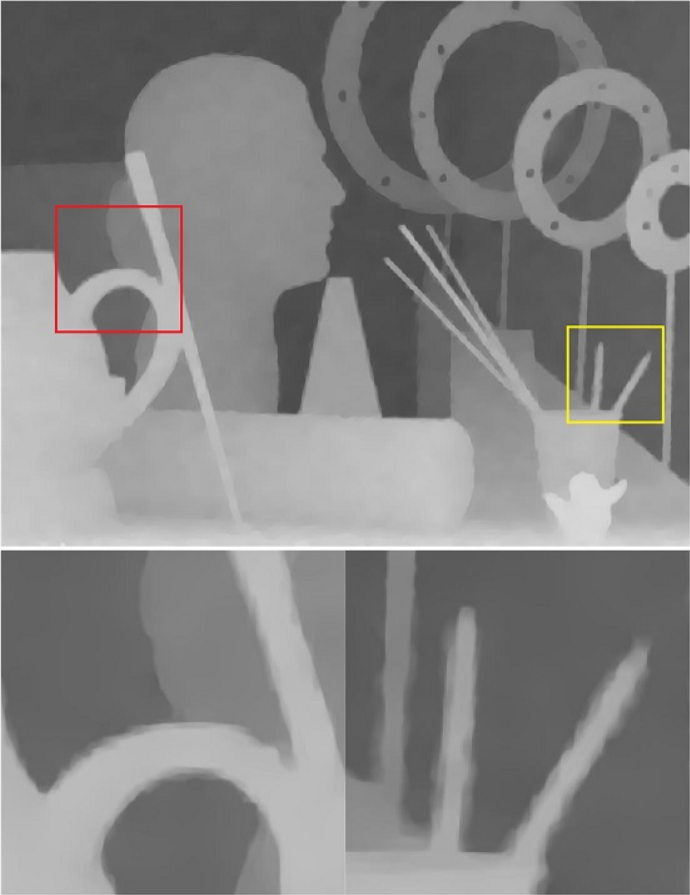}
  }
  \subfigure[]
  {
  \includegraphics[width=0.122\linewidth]{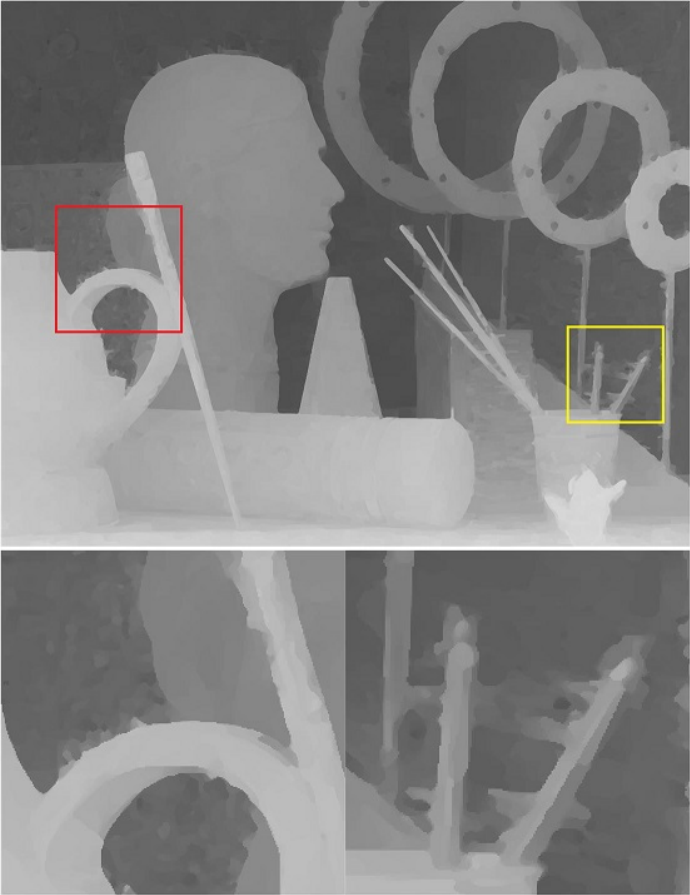}
  }
  \subfigure[]
  {
  \includegraphics[width=0.122\linewidth]{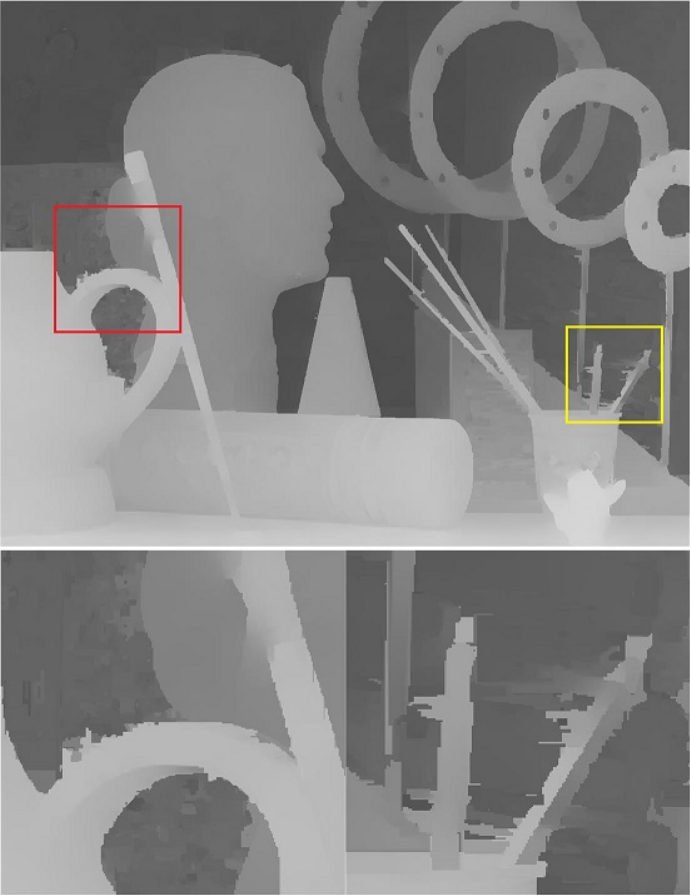}
  }
  \subfigure[]
  {
  \includegraphics[width=0.122\linewidth]{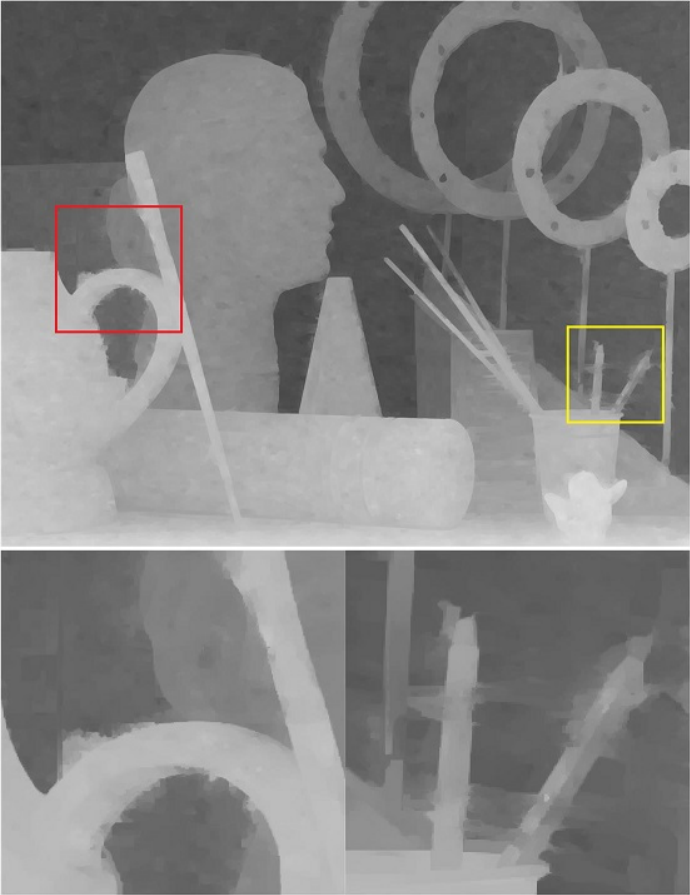}
  }
  \subfigure[]
  {
  \includegraphics[width=0.122\linewidth]{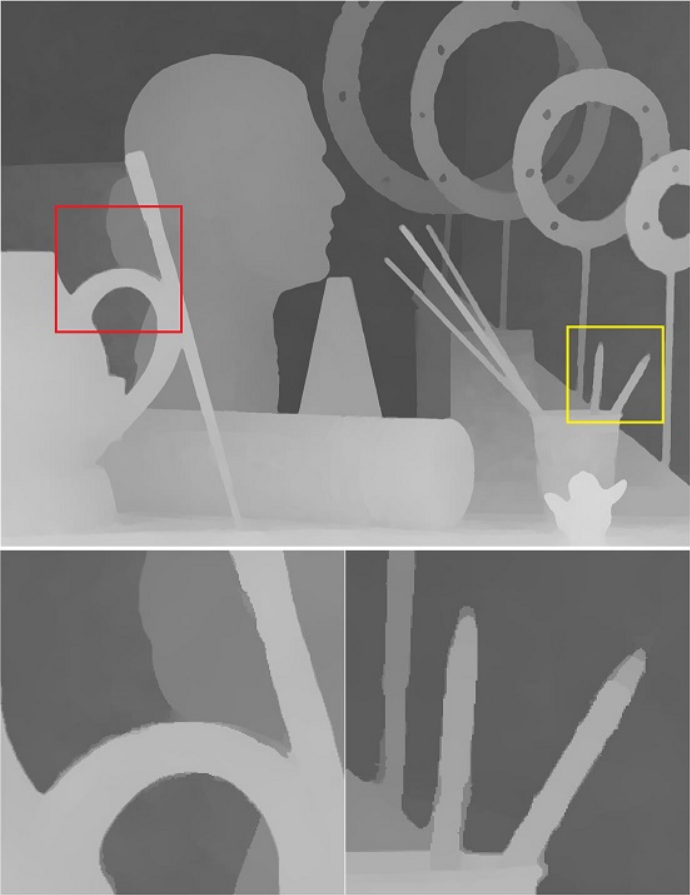}
  }
  \caption{Guided depth map upsampling results of simulated ToF data. (a) Guidance color image. (b) Ground-truth depth map. Result of (c) the approach proposed by Gu et~al. \cite{gu2017learning}, (d) SGF \cite{zhang2015segment}, (e) SD filter \cite{ham2015robust}, (f) Park et~al. \cite{park2011high} and (g) our method.}\label{FigToFSimulated}
\end{figure*}

\begin{figure*}
  \centering
  \subfigure[]
  {
  \includegraphics[width=0.122\linewidth]{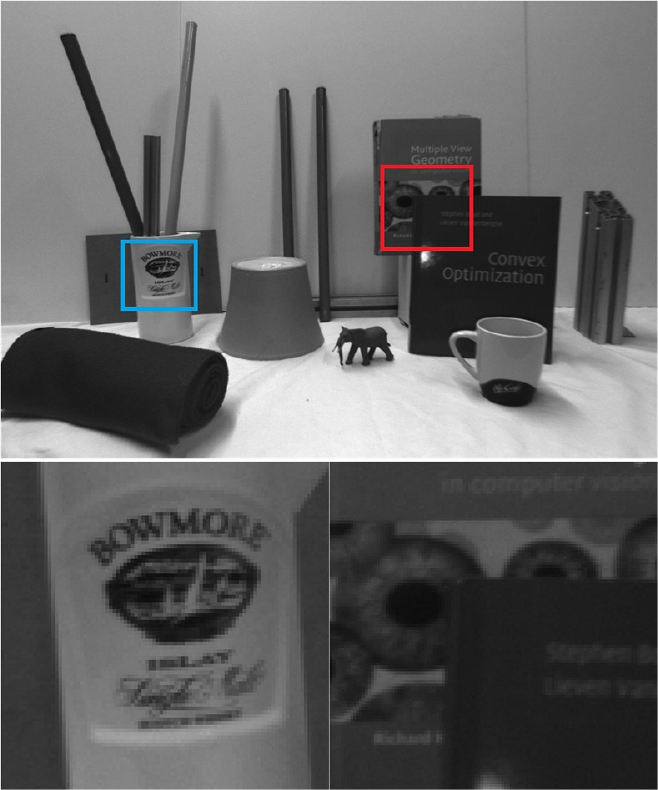}
  }
  \subfigure[]
  {
  \includegraphics[width=0.122\linewidth]{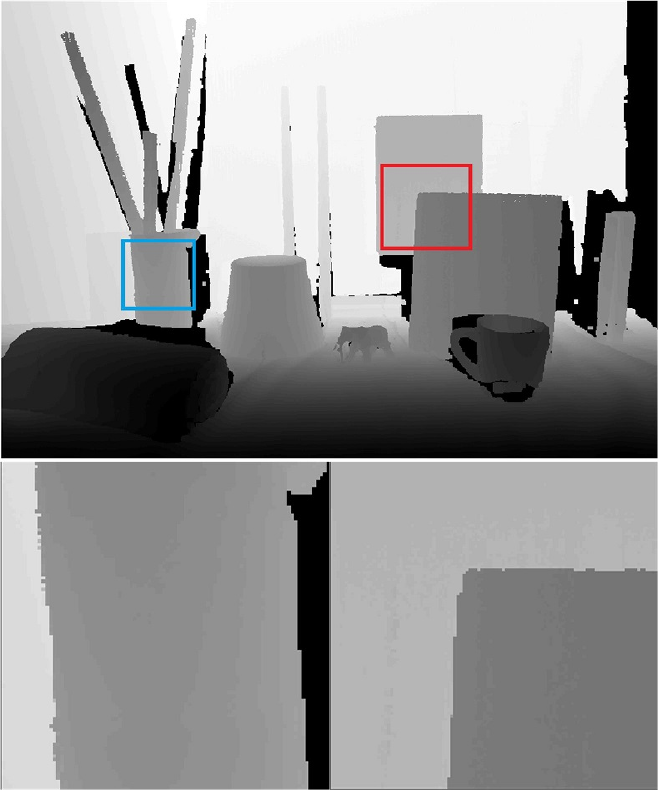}
  }
  \subfigure[]
  {
  \includegraphics[width=0.122\linewidth]{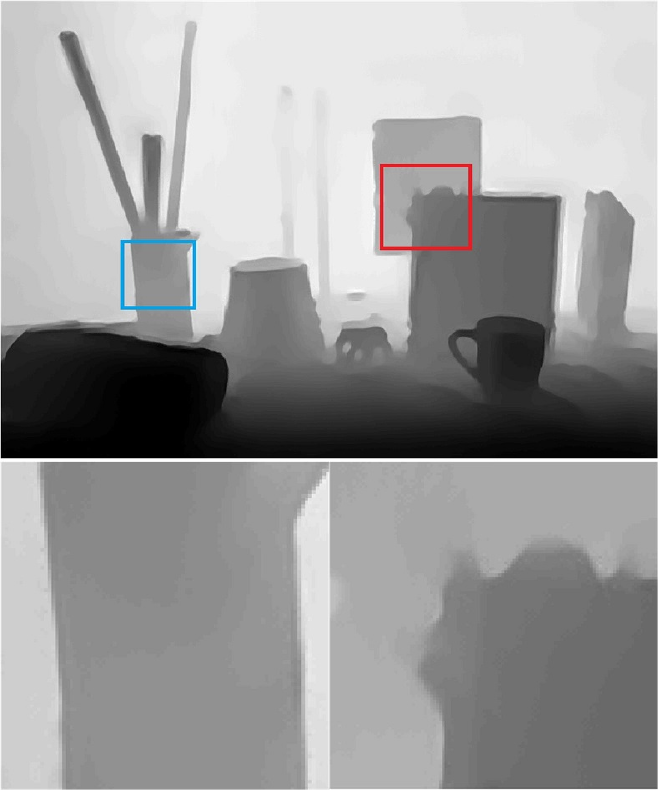}
  }
  \subfigure[]
  {
  \includegraphics[width=0.122\linewidth]{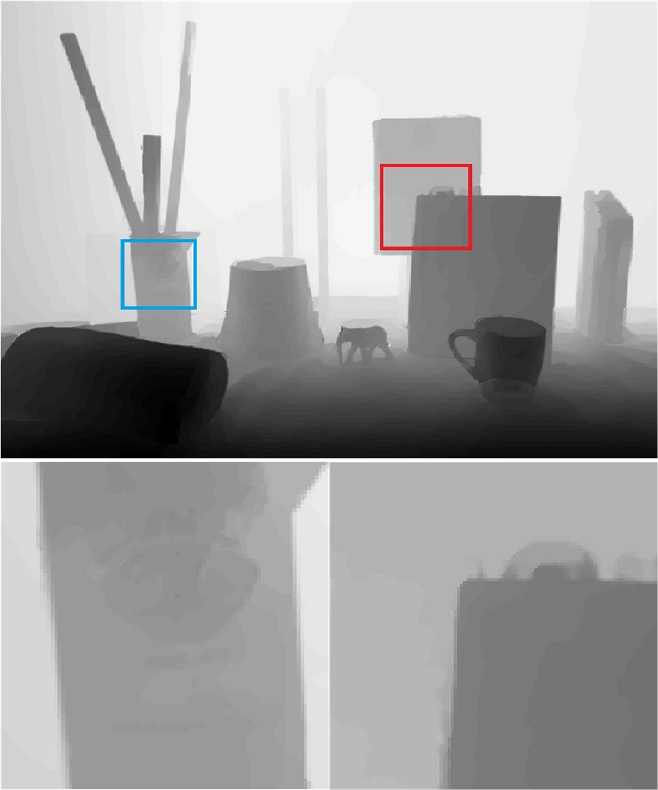}
  }
  \subfigure[]
  {
  \includegraphics[width=0.122\linewidth]{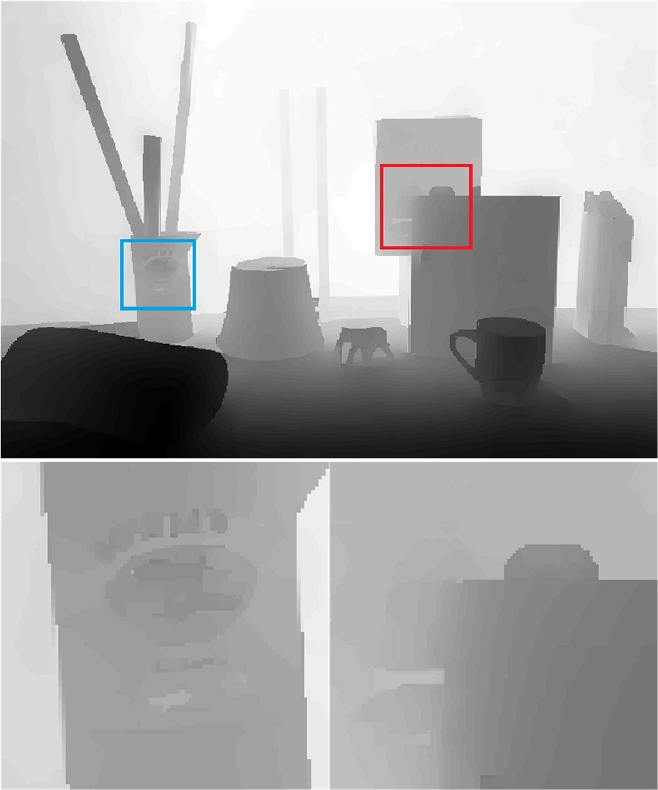}
  }
  \subfigure[]
  {
  \includegraphics[width=0.122\linewidth]{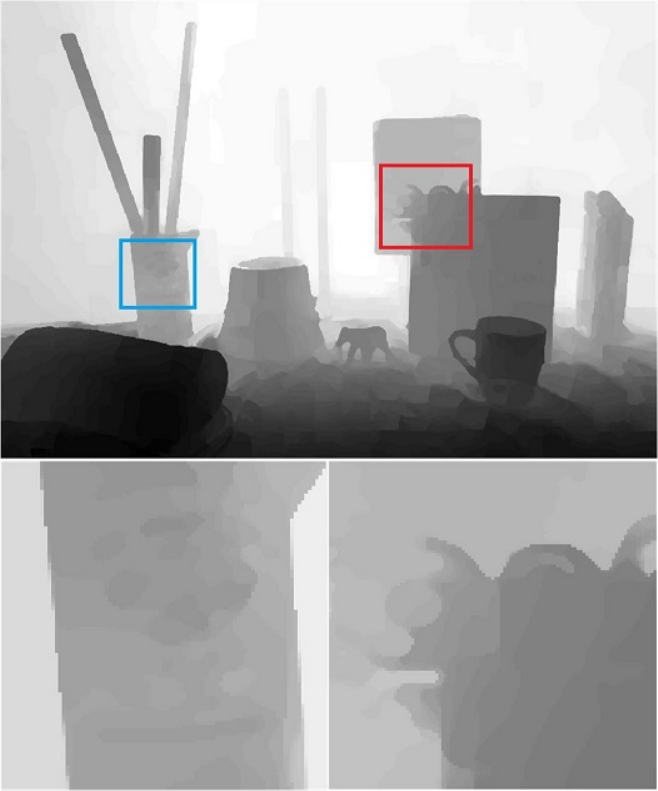}
  }
  \subfigure[]
  {
  \includegraphics[width=0.122\linewidth]{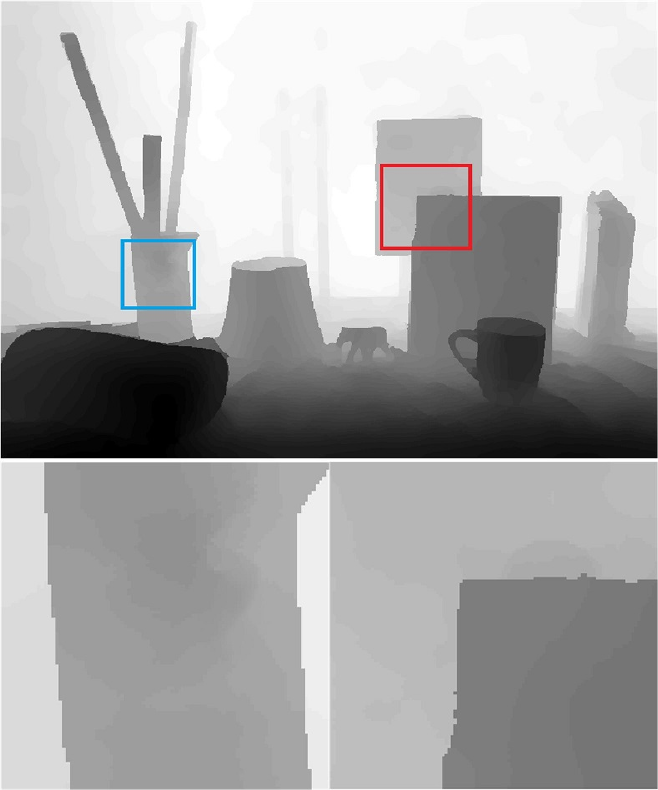}
  }
  \caption{Guided depth upsampling results of real ToF data. (a) Guidance intensity image. (b) Ground-truth depth map. Result of (c) the approach proposed by Gu et~al. \cite{gu2017learning}, (d) TGV \cite{ferstl2013image}, (e) SD filter \cite{ham2015robust}, (f) SGF \cite{zhang2015segment} and (g) our method.}\label{FigToFReal}
\end{figure*}

\begin{figure*}[!t]
  \centering
  \subfigure[]
  {
  \includegraphics[width=0.122\linewidth]{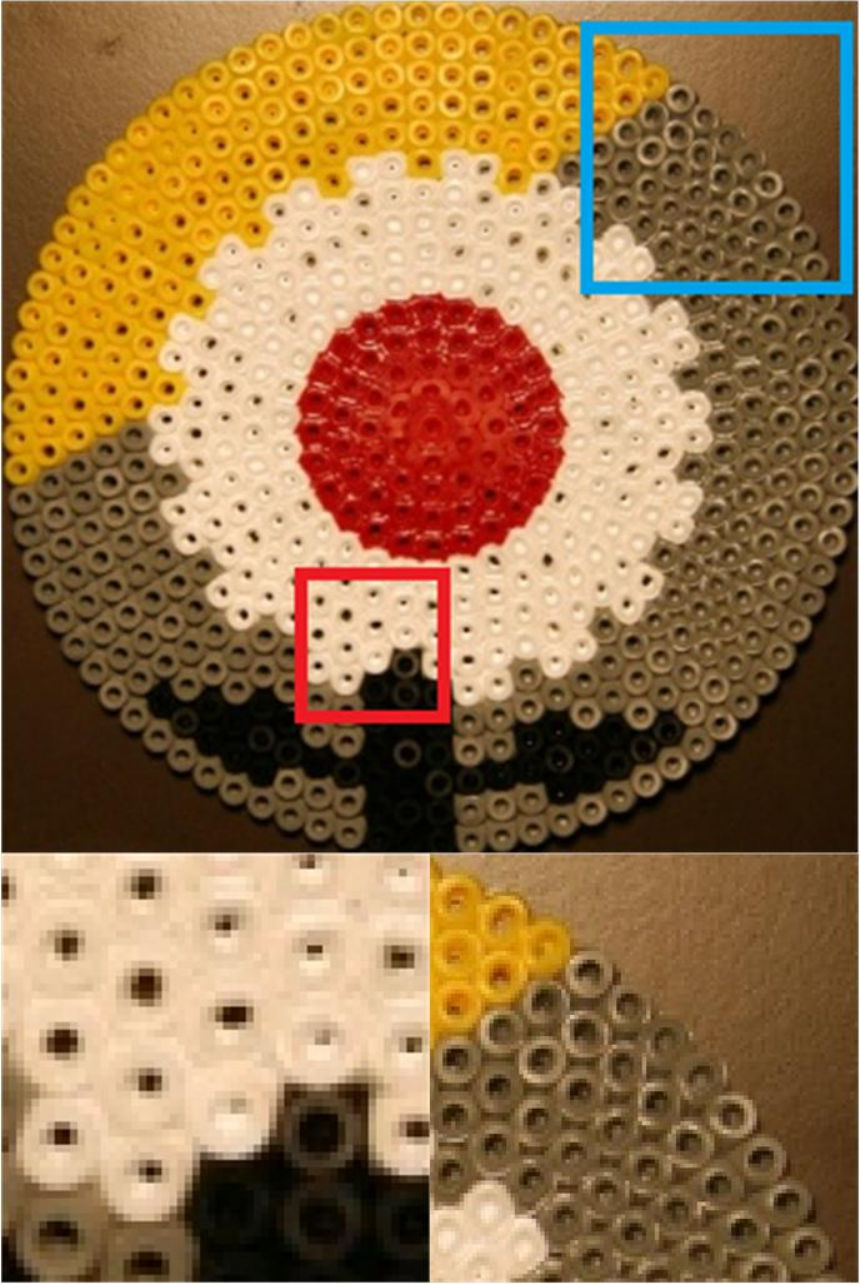}
  }
  \subfigure[]
  {
  \includegraphics[width=0.122\linewidth]{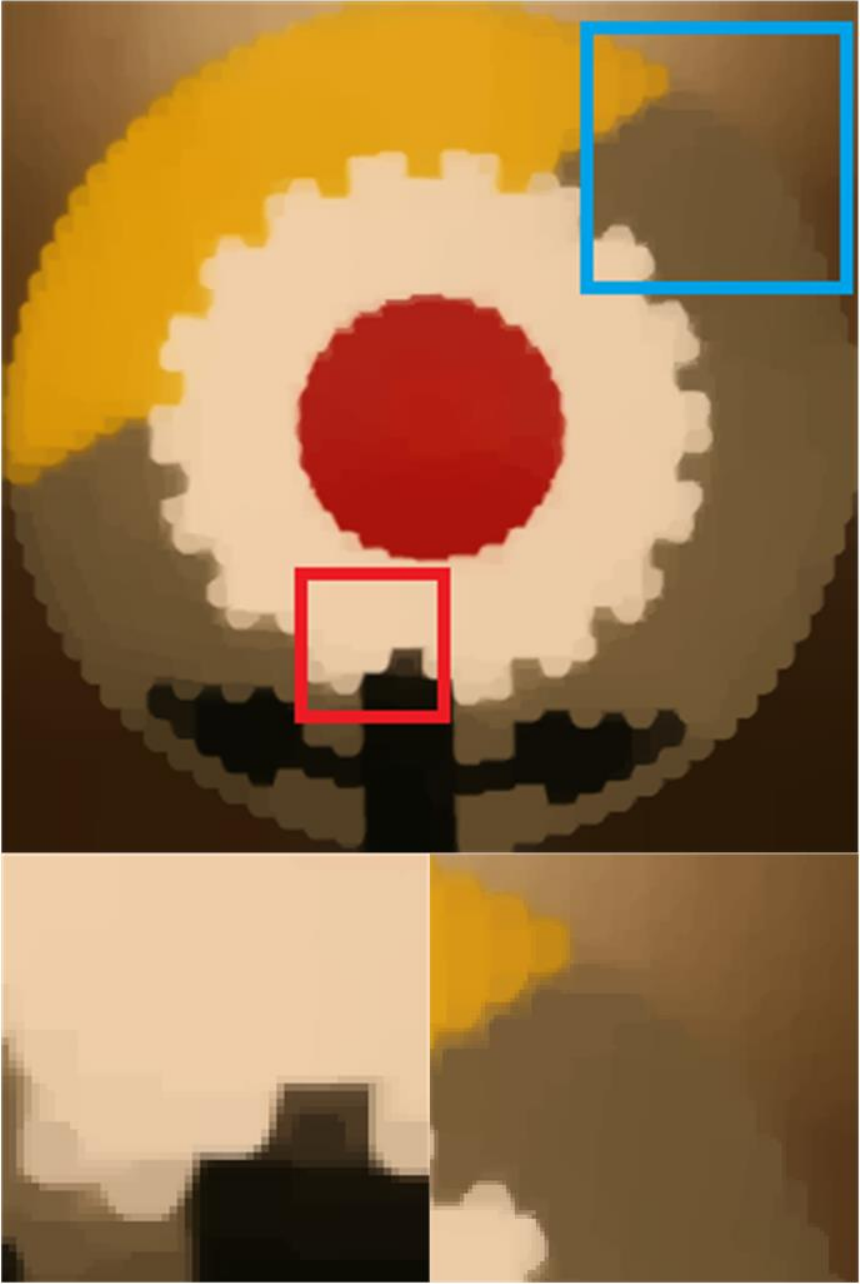}
  }
  \subfigure[]
  {
  \includegraphics[width=0.122\linewidth]{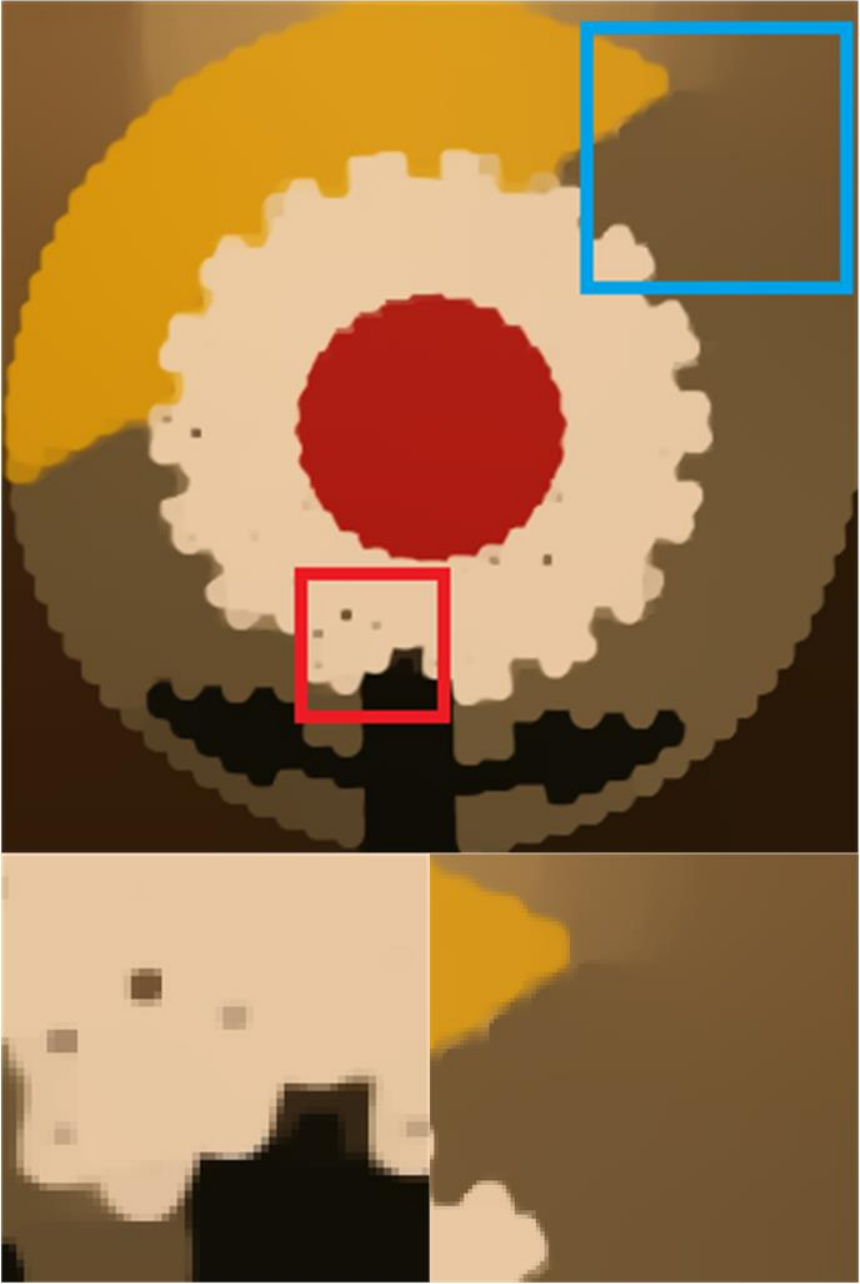}
  }
  \subfigure[]
  {
  \includegraphics[width=0.122\linewidth]{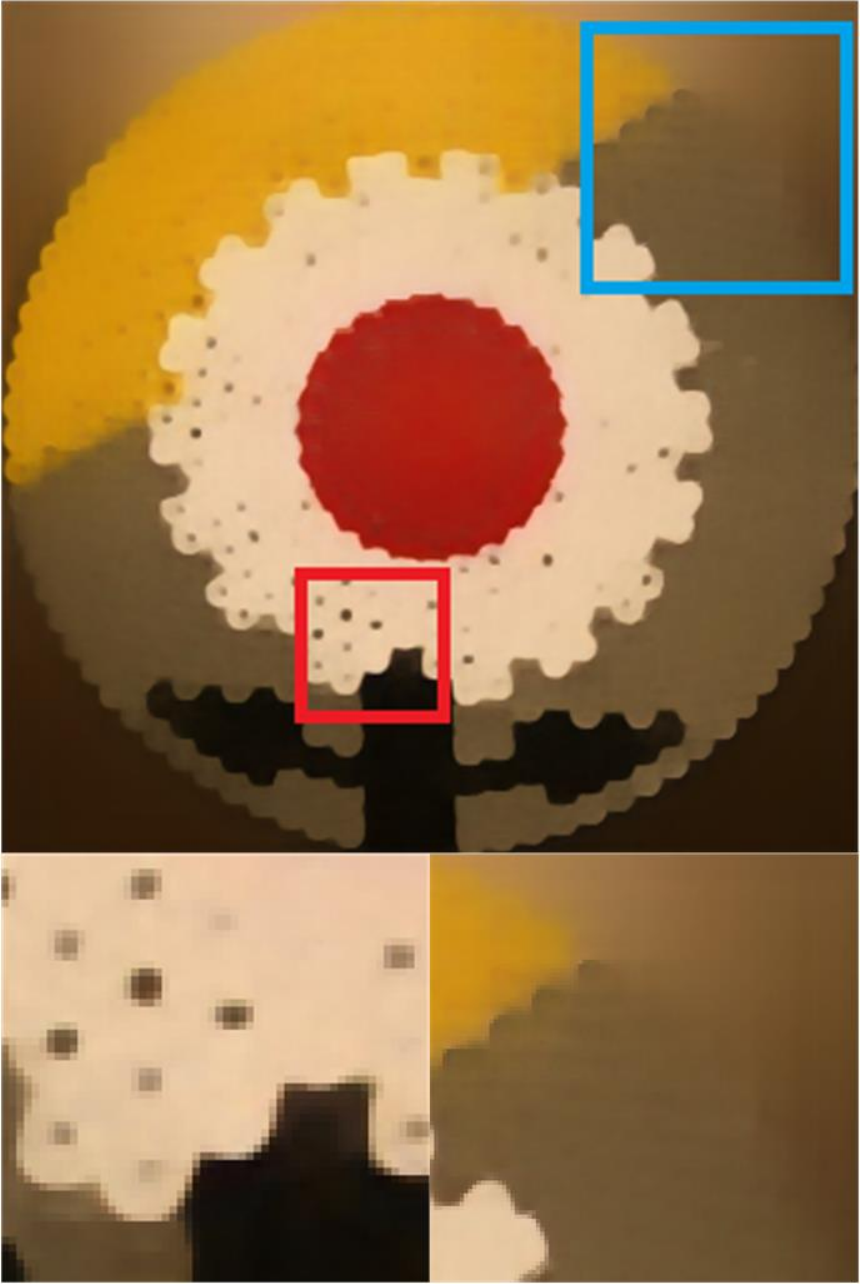}
  }
  \subfigure[]
  {
  \includegraphics[width=0.122\linewidth]{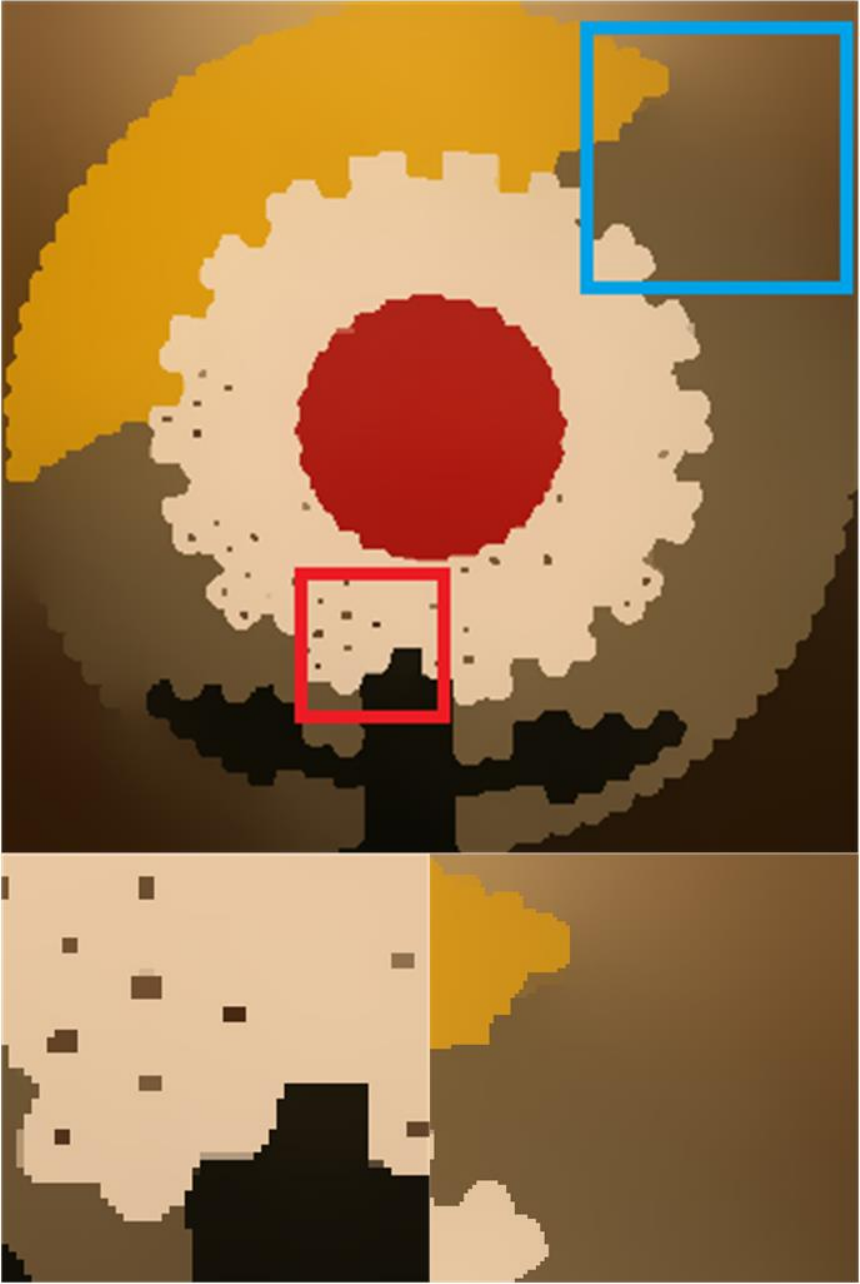}
  }
  \subfigure[]
  {
  \includegraphics[width=0.122\linewidth]{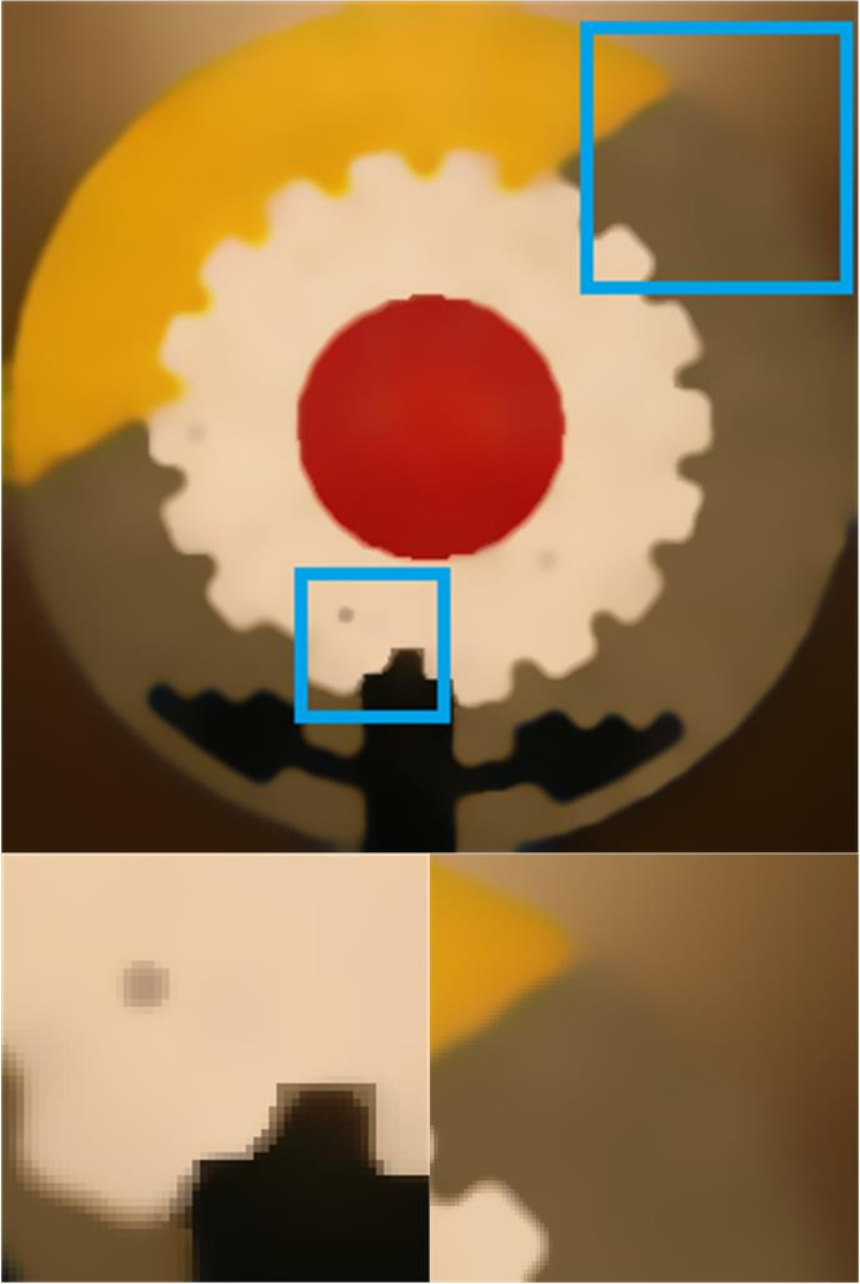}
  }
  \subfigure[]
  {
  \includegraphics[width=0.122\linewidth]{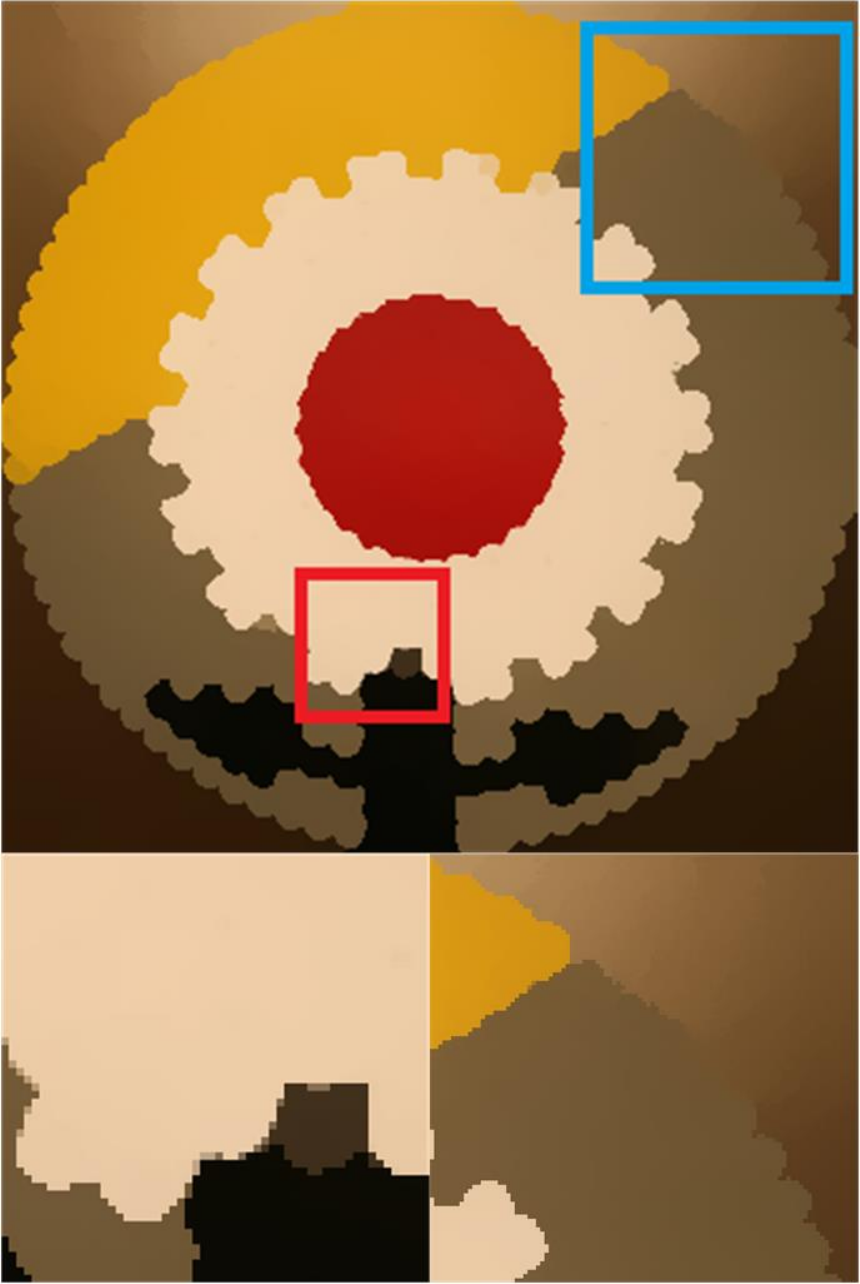}
  }
  \caption{Image texture removal results. (a) Input image. Result of (b) JCAS \cite{gu2017joint}, (c) RTV \cite{xu2012structure}, (d) FCN based approach \cite{chen2017fast}, (e) muGIF \cite{guo2018mutually} (f) BTF \cite{cho2014bilateral} and (g) our method.}\label{FigTextureSmooth}
\end{figure*}

The structure inconsistency issue in the third group can also be easily handled by our model. Note that $\mu_{i,j}^s$ in Eq.~(\ref{EqObjFunAuxULMu}) is computed with the smoothed image in each iteration, as formulated in Eq.~(\ref{EqMultHQCondition}), it thus can reflect the inherent natures of the smoothed image. The guidance weight $\omega_{i,j}$ can provide additional structural information from the guidance image $g$. This means that $\mu_{i,j}^s$ and $\omega_{i,j}$ can complement each other. In fact, the equivalent guidance weight of Eq.~(\ref{EqObjFunAuxULMu}) in each iteration is $\mu_{i,j}^s\omega_{i,j}$, which can reflect the property of both the smoothed image and the guidance image. In this way, it can properly handle the structure inconsistency problem and avoid blurring edges and texture copy artifacts. Similar ideas were also adopted in \cite{ham2015robust,liu2017robust}.

\section{Applications and Experimental Results}
\label{SecExperiments}

Our method is applied to various tasks in the first to the fourth groups to validate the effectiveness. Comparisons with the state-of-the-art approaches in each application are also presented. Due to the limited space, we only show experimental results of four applications. 

Our experiments are performed on a PC with an Intel Core i5 3.4GHz CPU (one thread used) and 8GB memory. For an RGB image of size $800\times600$ and $N=10$ in Algorithm \ref{Alg}, the running time is $10.04/25.09/43.11/69.82/96.73$ seconds in MATLAB for $r_d=r_s=1/2/3/4/5$. Note that as described in the property analysis section, the value of $r_d=r_s$ is smaller than 3 in most cases except for guided depth map upsampling. For the tasks in the first group which require $N=1$, the computational cost could be further reduced to $\frac{1}{10}$ of that mentioned above.

\textbf{HDR tone mapping} is a representative task in the first group. It requires to decompose the input image into a base layer and a detail layer through edge-preserving smoothing. The challenge of this task is that if the edges are sharpened by the smoothing procedure, it will result in gradient reversals, and halos will occur if the edges are blurred. Fig.~\ref{FigHDRToneMapping} shows the tone mapping results using different edge-preserving smoothing operators. The results of BF \cite{tomasi1998bilateral} and GF \cite{he2013guided} contain clear halos around the picture frames and the light fixture, as shown in Fig.~\ref{FigHDRToneMapping}(a) and (b). This is due to their local smoothing natures where strong smoothing can also blur salient edges \cite{farbman2008edge,he2013guided}. The $L_0$ norm smoothing \cite{xu2011image} can properly eliminate halos, but there are gradient reversals in its result as illustrated in Fig.~\ref{FigHDRToneMapping}(c). This is because the $L_0$ smoothing is prone to sharpen salient edges. The WLS \cite{farbman2008edge} and SG-WLS \cite{liu2017semi} smoothing perform well in handling gradient reversals and halos in most cases. However, there are slight halos in their results as illustrated in the left close-up in Fig.~\ref{FigHDRToneMapping}(d) and (e). These artifacts are properly eliminated in our results.

\begin{table*}
\centering
\caption{Quantitative comparison on the noisy simulated ToF data. Results are evaluated in MAE. The best results are in \textbf{bold}. The second best results are \underline{underlined}.}\label{TabToFSimulated}

\resizebox{1\textwidth}{!}
{
\begin{tabular}{|c|cccc|cccc|cccc|cccc|cccc|cccc|}

\hline
  \multicolumn{1}{|c}{\multirow{2}{*}{}} & \multicolumn{4}{|c|}{\emph{Art}} & \multicolumn{4}{c|}{\emph{Book}} & \multicolumn{4}{c|}{\emph{Dolls}} & \multicolumn{4}{c|}{\emph{Laundry} } & \multicolumn{4}{c|}{\emph{Moebius}} & \multicolumn{4}{c|}{\emph{Reindeer}}\\

  \cline{2-25} 
  & $2\times$ & $4\times$ & $8\times$ & $16\times$ & $2\times$ & $4\times$ & $8\times$ & $16\times$ & $2\times$ & $4\times$ & $8\times$ & $16\times$ & $2\times$ & $4\times$ & $8\times$ & $16\times$ & $2\times$ & $4\times$ & $8\times$ & $16\times$ & $2\times$ & $4\times$ & $8\times$ & $16\times$ \\
  \hline

  TGV\cite{ferstl2013image} & 0.8 & \underline{1.21} & 2.01 & 4.59 & 0.61 & 0.88 & 1.21 & 2.19 & 0.66 & \underline{0.95} & \underline{1.38} & 2.88 & \underline{0.61} & \textbf{0.87} & \underline{1.36} & 3.06 & {0.57} & \underline{0.77} & 1.23 & 2.74 & 0.61 & \textbf{0.85} & \underline{1.3} & 3.41 \\

  AR\cite{yang2014color} & 1.17 & 1.7 & 2.93 & 5.32 & 0.98 & 1.22 & 1.74 & 2.89 & 0.97 & 1.21 & 1.71 & 2.74 & 1 & 1.31 & 1.97 & 3.43 & 0.95 & 1.2 & 1.79 & 2.82 & 1.07 & 1.3 & 2.03 & 3.34\\

  SG-WLS\cite{liu2017semi} & 1.26 & 1.9 & 3.07 & - & 0.82 & 1.12 & 1.73 & - & 0.87 & 1.11 &	1.81 & - & 0.86 & 1.17 & 2 & - & 0.82 & 1.08 &1.79 & - & 0.9 & 1.32 & 2.01 & -\\

  FGI\cite{li2016fast} & 0.9 & 1.37 & 2.46 & 4.89 & 0.66 & \underline{0.85} & \underline{1.23} & \underline{1.96} & 0.74 & \underline{0.95} & 1.41 & \underline{2.13} & 0.71 & 0.99 & 1.59 & 2.67 & 0.67 & 0.82 & \underline{1.2} & \underline{1.87} & 0.75 & 0.94 & 1.55 & 2.73\\

  SGF\cite{zhang2015segment} & 1.42 & 1.85 & 3.06 & 5.55 & 0.84 & 1.11 & 1.76 & 3.03 & 0.87 & 1.2 & 1.88 & 3.26 & 0.74 & 1.1 & 1.96 & 3.63 & 0.81 & 1.13 & 1.84 & 3.16 & 0.93 & 1.25 & 2.03 & 3.67\\

  SD Filter\cite{ham2015robust} & 1.16 & 1.64 & 2.88 & 5.52 & 0.86 & 1.1 & 1.57 & 2.68 & 1.04 & 1.27 & 1.73 & 2.76 & 0.96 & 1.25 & 1.94 & 3.54 & 0.93 & 1.14 & 1.68 & 2.75 & 1.05 & 1.31 & 1.99 & 3.43\\

  FBS\cite{barron2016fast} & 1.93 & 2.39 & 3.29 & 5.05 & 1.42 & 1.55 & 1.76 & 2.48 & 1.33 & 1.45 & 1.69 & 2.26 & 1.32 & 1.49 & 1.77 & 2.67 & 1.16 & 1.29 & 1.61 & 2.44 & 1.63 & 1.76 & 2.01 & 2.69\\

  muGIF\cite{guo2018mutually} & 1.00 & 1.26 & \underline{2.00} & \underline{3.46} & 0.73 & 0.89 & 1.35 & 2.15 & 0.85 & 1.04 & 1.50 & 2.45 & 0.64 & \textbf{0.87} & \underline{1.36} & \underline{2.57} & 0.67 & 0.85 & 1.35 & 2.25 & 0.78 & 0.94 & 1.39 & \underline{2.52}\\

  Park et~al.\cite{park2011high} & 1.66 & 2.47 & 3.44 & 5.55 & 1.19 & 1.47 & 2.06 & 3.1 & 1.19 & 1.56 & 2.15 & 3.04 & 1.34 & 1.73 & 2.41 & 3.85 & 1.2 & 1.5 & 2.13 & 2.95 & 1.26 & 1.65 & 2.46 & 3.66 \\

  Shen et~al.\cite{shen2015mutual} & 1.79 & 2.21 & 3.2 & 5.04 & 1.34 & 1.69 & 2.25 & 3.13 & 1.37 & 1.58 & 2.05 & 2.85 & 1.49 & 1.74 & 2.34 & 3.5 & 1.34 & 1.56 & 2.09 & 2.99 & 1.29 & 1.55 & 2.19 & 3.33\\

  Gu et~al.\cite{gu2017learning} & \textbf{0.61} & 1.46 & 2.98 & 5.09 & \textbf{0.52} & 0.95 & 1.87 & 2.98 & \underline{0.63} & 1.02 & 1.89 & 2.92 & \textbf{0.58} & 1.14 & 2.21 & 3.58 & \underline{0.53} & 0.96 & 1.89 & 2.99 & \textbf{0.52} & 1.07 & 2.17 & 3.59\\

  Li et~al.\cite{li2016deep} & - & 3.77 & 4.49 & 6.29 & - & 3.21 & 3.28 & 3.79 & - & 3.19 & 3.28 & 3.79 & - & 3.34 & 3.61 & 4.45 & - & 3.23 & 3.35 & 3.92 & - & 3.39 & 3.65 & 4.54\\

  Ours & \underline{0.69} & \textbf{1.07} & \textbf{1.65} & \textbf{2.96} & \underline{0.55} & \textbf{0.81} & \textbf{1.22} & \textbf{1.78} & \textbf{0.62} & \textbf{0.9} & \textbf{1.27} & \textbf{1.84} & \underline{0.61} & \underline{0.89} & \textbf{1.28} & \textbf{2.12} & \textbf{0.51} & \textbf{0.75} & \textbf{1.12} & \textbf{1.71} & \underline{0.56} & \underline{0.87} & \textbf{1.27} & \textbf{2.08}\\
  \hline
\end{tabular}
}\vspace{-0.5em}
\end{table*}

\begin{table}
\centering
\caption{Quantitative comparison on real ToF dataset. The errors are calculated as MAE to the measured ground-truth in \texttt{mm}. The best results are in \textbf{bold}. The second best results are \underline{underlined}.}\label{TabToFReal}

\resizebox{1\linewidth}{!}
{
\begin{tabular}{|c|ccc|}
  \hline
   & \emph{Books} & \emph{Devil} & \emph{Shark}\\
  \hline
  Bicubic & 16.23mm & 17.78mm & 16.66mm\\
  GF\cite{he2013guided} & 15.55mm & 16.1mm & 17.1mm\\
  SD Filter\cite{ham2015robust} & 13.47mm & 15.99mm & 16.18mm\\
  SG-WLS\cite{liu2017semi} & 14.71mm & 16.24mm & 16.51mm\\
  Shen et~al.\cite{shen2015mutual}  & 15.47mm & 16.18mm & 17.33mm\\
  Park et~al.\cite{park2011high} & 14.31mm & 15.36mm & 15.88mm\\
  TGV\cite{ferstl2013image} &\underline{12.8mm} &\underline{14.97mm} & \underline{15.53mm}\\
  AR\cite{yang2014color} & 14.37mm & 15.41mm & 16.27mm\\
  Gu et~al.\cite{gu2017learning} & 13.87mm & 15.36mm & 15.88mm\\
  SGF\cite{zhang2015segment} & 13.57mm & 15.74mm & 16.21mm\\
  FGI\cite{li2016fast} & 14.21mm & 16.43mm & 16.37mm\\
  FBS\cite{barron2016fast} & 15.93mm & 17.21mm & 16.33mm\\
  Li et~al.\cite{li2016deep} & 14.33mm & 15.09mm & 15.82mm\\
  Ours & \textbf{12.49mm} & \textbf{14.51mm} & \textbf{15.02mm}\\
  \hline
\end{tabular}
}
\end{table}

\textbf{Clip-art compression artifacts removal}. Clip-art images are piecewise constant with sharp edges. When they are compressed in JPEG format with low quality, there will be edge-related artifacts, and the edges are usually blurred as shown in Fig.~\ref{FigClipArt}(a). Therefore, when removing the compression artifacts, the edges should also be sharpened in the restored image. We thus classify this task into the second group. The approach proposed by Wang et~al. \cite{wang2006deringing} can seldom handle heavy compression artifacts as shown in Fig.~\ref{FigClipArt}(b). The $L_0$ norm smoothing fails to preserve weak edges as shown in Fig.~\ref{FigClipArt}(c). The region fusion approach \cite{nguyen2015fast} is able to produce results with sharpened edges, however, it also enhances the blocky artifacts along strong edges as highlighted in Fig.~\ref{FigClipArt}(d). The edges in the result of BTF \cite{cho2014bilateral} are blurred in Fig.~\ref{FigClipArt}(e).  Our result is illustrated in Fig.~\ref{FigClipArt}(f) with edges sharpened and compression artifacts removed.

\textbf{Guided depth map upsampling} belongs to the guided image filtering in the third group. The RGB guided image can provide additional structural information to restore and sharpen the depth edges. The challenge of this task is the structure inconsistency between the depth map and the RGB guidance image, which can cause blurring depth edges and texture copy artifacts in the upsampled depth map. We test our method on the simulated dateset provided in \cite{yang2014color}. Fig.~\ref{FigToFSimulated} shows the visual comparison between our result and the results of the recent state-of-the-art approaches. Our method shows better performance in preserving sharp depth edges and avoiding texture copy artifacts. Tab.~\ref{TabToFSimulated} also shows the quantitative evaluation on the results of different methods. Following the measurement used in \cite{guo2018mutually,li2016fast,liu2017semi,yang2014color}, the evaluation is measured in terms of mean absolute errors (MAE). As Tab.~\ref{TabToFSimulated} shows, our method can achieve the best or the second best performance among all the compared approaches.

We further validate our method on the real data introduced by Ferstl et~al. \cite{ferstl2013image}. The real dataset contains three low-resolution depth maps captured by a ToF depth camera and the corresponding highly accurate ground-truth depth maps captured with structured light. The upsampling factor for the real dataset is $\sim6.25\times$. The visual comparison in Fig.~\ref{FigToFReal} and the quantitative comparison in Tab.~\ref{TabToFReal} shows that our method can outperform the compared methods and achieve state-of-the-art performance.

\textbf{Image texture removal} belongs to the tasks in the fourth group. It aims at extracting salient meaningful structures while removing small complex texture patterns. The challenge of this task is that it requires structure-preserving smoothing rather than the edge-preserving in the above tasks. Fig.~\ref{FigTextureSmooth}(a) shows a classical example of image texture removal: the small textures with strong edges should be smoothed out while the salient structures with weak edges should be preserved. Fig.~\ref{FigTextureSmooth}(b)$\sim$(f) show the results of the recent state-of-the-art approaches. The joint convolutional analysis and synthesis sparse (JCAS) model \cite{gu2017joint} can well remove the textures, but the resulting edges are also blurred. The RTV method \cite{xu2012structure}, muGIF \cite{guo2018mutually}, BTF \cite{cho2014bilateral} and FCN based approach \cite{chen2017fast} cannot completely remove the textures, in addition, the weak edges of the salient structures have also been smoothed out in their results. Our method can both preserve the weak edges of the salient structures and remove the small textures.

\section{Conclusion}

We propose a non-convex non-smooth optimization framework for edge-preserving and structure-preserving image smoothing. We first introduce the truncated Huber penalty function which shows strong flexibility. Then a robust framework is presented. When combined with the flexibility of the truncated Huber penalty function, our framework is able to achieve different and even contradictive smoothing behaviors using different parameter settings. This is different from most previous approaches of which the inherent smoothing natures are usually fixed. We further propose an efficient numerical solution to our model and prove its convergence theoretically. Comprehensive experimental results in a number of applications demonstrate the effectiveness of our method.

\noindent\textbf{Acknowledgement}\\
We gratefully acknowledge the support of the Australia Centre for Robotic Vision. This paper is also partly supported by NSFC, China (No. U1803261, 61977046), Key Research and Development Program of Sichuan Province (No. 2019YFG0409) and National Key Research and Development Project (No. 2018AAA0100702)

{\small
\bibliographystyle{aaai}
\bibliography{egbib}

\begin{thebibliography}{}

\bibitem[\protect\citeauthoryear{Barron and Poole}{2016}]{barron2016fast}
Barron, J.~T., and Poole, B.
\newblock 2016.
\newblock The fast bilateral solver.
\newblock In {\em ECCV},  617--632.
\newblock Springer.

\bibitem[\protect\citeauthoryear{Buades \bgroup et al\mbox.\egroup
  }{2010}]{buades2010fast}
Buades, A.; Le, T.~M.; Morel, J.-M.; Vese, L.~A.; et~al.
\newblock 2010.
\newblock Fast cartoon+ texture image filters.
\newblock {\em TIP} 19(8):1978--1986.

\bibitem[\protect\citeauthoryear{Chan and Esedoglu}{2005}]{chan2005aspects}
Chan, T.~F., and Esedoglu, S.
\newblock 2005.
\newblock Aspects of total variation regularized l 1 function approximation.
\newblock {\em SIAM Journal on Applied Mathematics} 65(5):1817--1837.

\bibitem[\protect\citeauthoryear{Chen, Xu, and Koltun}{2017}]{chen2017fast}
Chen, Q.; Xu, J.; and Koltun, V.
\newblock 2017.
\newblock Fast image processing with fully-convolutional networks.
\newblock In {\em ICCV}, volume~9,  2516--2525.

\bibitem[\protect\citeauthoryear{Cho \bgroup et al\mbox.\egroup
  }{2014}]{cho2014bilateral}
Cho, H.; Lee, H.; Kang, H.; and Lee, S.
\newblock 2014.
\newblock Bilateral texture filtering.
\newblock {\em ToG} 33(4):128.

\bibitem[\protect\citeauthoryear{Durand and Dorsey}{2002}]{durand2002fast}
Durand, F., and Dorsey, J.
\newblock 2002.
\newblock Fast bilateral filtering for the display of high-dynamic-range
  images.
\newblock In {\em ToG}, volume~21,  257--266.
\newblock ACM.

\bibitem[\protect\citeauthoryear{Fan \bgroup et al\mbox.\egroup
  }{2018}]{fan2018image}
Fan, Q.; Yang, J.; Wipf, D.; Chen, B.; and Tong, X.
\newblock 2018.
\newblock Image smoothing via unsupervised learning.
\newblock In {\em SIGGRAPH Asia 2018 Technical Papers},  259.
\newblock ACM.

\bibitem[\protect\citeauthoryear{Fan \bgroup et al\mbox.\egroup
  }{2019}]{fan2019general}
Fan, Q.; Chen, D.; Yuan, L.; Hua, G.; Yu, N.; and Chen, B.
\newblock 2019.
\newblock A general decoupled learning framework for parameterized image
  operators.
\newblock {\em IEEE transactions on pattern analysis and machine intelligence}.

\bibitem[\protect\citeauthoryear{Farbman \bgroup et al\mbox.\egroup
  }{2008}]{farbman2008edge}
Farbman, Z.; Fattal, R.; Lischinski, D.; and Szeliski, R.
\newblock 2008.
\newblock Edge-preserving decompositions for multi-scale tone and detail
  manipulation.
\newblock In {\em ToG}, volume~27, ~67.
\newblock ACM.

\bibitem[\protect\citeauthoryear{Fattal, Agrawala, and
  Rusinkiewicz}{2007}]{fattal2007multiscale}
Fattal, R.; Agrawala, M.; and Rusinkiewicz, S.
\newblock 2007.
\newblock Multiscale shape and detail enhancement from multi-light image
  collections.
\newblock In {\em ToG}, volume~26, ~51.
\newblock ACM.

\bibitem[\protect\citeauthoryear{Ferstl \bgroup et al\mbox.\egroup
  }{2013}]{ferstl2013image}
Ferstl, D.; Reinbacher, C.; Ranftl, R.; R{\"u}ther, M.; and Bischof, H.
\newblock 2013.
\newblock Image guided depth upsampling using anisotropic total generalized
  variation.
\newblock In {\em ICCV},  993--1000.

\bibitem[\protect\citeauthoryear{Gastal and Oliveira}{2011}]{gastal2011domain}
Gastal, E.~S., and Oliveira, M.~M.
\newblock 2011.
\newblock Domain transform for edge-aware image and video processing.
\newblock In {\em ToG}, volume~30, ~69.
\newblock ACM.

\bibitem[\protect\citeauthoryear{Gastal and
  Oliveira}{2012}]{gastal2012adaptive}
Gastal, E.~S., and Oliveira, M.~M.
\newblock 2012.
\newblock Adaptive manifolds for real-time high-dimensional filtering.
\newblock {\em ToG} 31(4):33.

\bibitem[\protect\citeauthoryear{Geman and Yang}{1995}]{geman1995nonlinear}
Geman, D., and Yang, C.
\newblock 1995.
\newblock Nonlinear image recovery with half-quadratic regularization.
\newblock {\em TIP} 4(7):932--946.

\bibitem[\protect\citeauthoryear{Gu \bgroup et al\mbox.\egroup
  }{2017a}]{gu2017joint}
Gu, S.; Meng, D.; Zuo, W.; and Zhang, L.
\newblock 2017a.
\newblock Joint convolutional analysis and synthesis sparse representation for
  single image layer separation.
\newblock In {\em ICCV},  1717--1725.
\newblock IEEE.

\bibitem[\protect\citeauthoryear{Gu \bgroup et al\mbox.\egroup
  }{2017b}]{gu2017learning}
Gu, S.; Zuo, W.; Guo, S.; Chen, Y.; Chen, C.; and Zhang, L.
\newblock 2017b.
\newblock Learning dynamic guidance for depth image enhancement.
\newblock In {\em CVPR}.

\bibitem[\protect\citeauthoryear{Guo \bgroup et al\mbox.\egroup
  }{2018}]{guo2018mutually}
Guo, X.; Li, Y.; Ma, J.; and Ling, H.
\newblock 2018.
\newblock Mutually guided image filtering.
\newblock {\em TPAMI}.

\bibitem[\protect\citeauthoryear{Ham, Cho, and Ponce}{2015}]{ham2015robust}
Ham, B.; Cho, M.; and Ponce, J.
\newblock 2015.
\newblock Robust image filtering using joint static and dynamic guidance.
\newblock In {\em CVPR},  4823--4831.

\bibitem[\protect\citeauthoryear{He, Sun, and Tang}{2013}]{he2013guided}
He, K.; Sun, J.; and Tang, X.
\newblock 2013.
\newblock Guided image filtering.
\newblock {\em TPAMI} 35(6):1397--1409.

\bibitem[\protect\citeauthoryear{Holland and Welsch}{1977}]{holland1977robust}
Holland, P.~W., and Welsch, R.~E.
\newblock 1977.
\newblock Robust regression using iteratively reweighted least-squares.
\newblock {\em Communications in Statistics-theory and Methods} 6(9):813--827.

\bibitem[\protect\citeauthoryear{Huber and others}{1964}]{huber1964robust}
Huber, P.~J., et~al.
\newblock 1964.
\newblock Robust estimation of a location parameter.
\newblock {\em The annals of mathematical statistics} 35(1):73--101.

\bibitem[\protect\citeauthoryear{Karacan, Erdem, and
  Erdem}{2013}]{karacan2013structure}
Karacan, L.; Erdem, E.; and Erdem, A.
\newblock 2013.
\newblock Structure-preserving image smoothing via region covariances.
\newblock {\em ToG} 32(6):176.

\bibitem[\protect\citeauthoryear{Kopf \bgroup et al\mbox.\egroup
  }{2007}]{kopf2007joint}
Kopf, J.; Cohen, M.~F.; Lischinski, D.; and Uyttendaele, M.
\newblock 2007.
\newblock Joint bilateral upsampling.
\newblock In {\em ToG}, volume~26, ~96.
\newblock ACM.

\bibitem[\protect\citeauthoryear{Lanckriet and
  Sriperumbudur}{2009}]{lanckriet2009convergence}
Lanckriet, G.~R., and Sriperumbudur, B.~K.
\newblock 2009.
\newblock On the convergence of the concave-convex procedure.
\newblock In {\em NeurIPS},  1759--1767.

\bibitem[\protect\citeauthoryear{Li \bgroup et al\mbox.\egroup
  }{2016a}]{li2016deep}
Li, Y.; Huang, J.-B.; Ahuja, N.; and Yang, M.-H.
\newblock 2016a.
\newblock Deep joint image filtering.
\newblock In {\em ECCV},  154--169.
\newblock Springer.

\bibitem[\protect\citeauthoryear{Li \bgroup et al\mbox.\egroup
  }{2016b}]{li2016fast}
Li, Y.; Min, D.; Do, M.~N.; and Lu, J.
\newblock 2016b.
\newblock Fast guided global interpolation for depth and motion.
\newblock In {\em ECCV},  717--733.
\newblock Springer.

\bibitem[\protect\citeauthoryear{Liu \bgroup et al\mbox.\egroup
  }{2017a}]{liu2017semi}
Liu, W.; Chen, X.; Shen, C.; Liu, Z.; and Yang, J.
\newblock 2017a.
\newblock Semi-global weighted least squares in image filtering.
\newblock In {\em ICCV},  5861--5869.

\bibitem[\protect\citeauthoryear{Liu \bgroup et al\mbox.\egroup
  }{2017b}]{liu2017robust}
Liu, W.; Chen, X.; Yang, J.; and Wu, Q.
\newblock 2017b.
\newblock Robust color guided depth map restoration.
\newblock {\em TIP} 26(1):315--327.

\bibitem[\protect\citeauthoryear{Nguyen and Brown}{2015}]{nguyen2015fast}
Nguyen, R.~M., and Brown, M.~S.
\newblock 2015.
\newblock Fast and effective l0 gradient minimization by region fusion.
\newblock In {\em ICCV},  208--216.

\bibitem[\protect\citeauthoryear{Nikolova and Ng}{2005}]{nikolova2005analysis}
Nikolova, M., and Ng, M.~K.
\newblock 2005.
\newblock Analysis of half-quadratic minimization methods for signal and image
  recovery.
\newblock {\em SIAM Journal on Scientific Computing} 27(3):937--966.

\bibitem[\protect\citeauthoryear{Nikolova}{2004}]{nikolova2004variational}
Nikolova, M.
\newblock 2004.
\newblock A variational approach to remove outliers and impulse noise.
\newblock {\em Journal of Mathematical Imaging and Vision} 20(1-2):99--120.

\bibitem[\protect\citeauthoryear{Park \bgroup et al\mbox.\egroup
  }{2011}]{park2011high}
Park, J.; Kim, H.; Tai, Y.-W.; Brown, M.~S.; and Kweon, I.
\newblock 2011.
\newblock High quality depth map upsampling for 3d-tof cameras.
\newblock In {\em ICCV},  1623--1630.
\newblock IEEE.

\bibitem[\protect\citeauthoryear{Petschnigg \bgroup et al\mbox.\egroup
  }{2004}]{petschnigg2004digital}
Petschnigg, G.; Szeliski, R.; Agrawala, M.; Cohen, M.; Hoppe, H.; and Toyama,
  K.
\newblock 2004.
\newblock Digital photography with flash and no-flash image pairs.
\newblock {\em ToG} 23(3):664--672.

\bibitem[\protect\citeauthoryear{Rudin, Osher, and
  Fatemi}{1992}]{rudin1992nonlinear}
Rudin, L.~I.; Osher, S.; and Fatemi, E.
\newblock 1992.
\newblock Nonlinear total variation based noise removal algorithms.
\newblock {\em Physica D: nonlinear phenomena} 60(1-4):259--268.

\bibitem[\protect\citeauthoryear{Shen \bgroup et al\mbox.\egroup
  }{2015a}]{shen2015multispectral}
Shen, X.; Yan, Q.; Xu, L.; Jia, J.; et~al.
\newblock 2015a.
\newblock Multispectral joint image restoration via optimizing a scale map.
\newblock {\em TPAMI} (1):1--1.

\bibitem[\protect\citeauthoryear{Shen \bgroup et al\mbox.\egroup
  }{2015b}]{shen2015mutual}
Shen, X.; Zhou, C.; Xu, L.; and Jia, J.
\newblock 2015b.
\newblock Mutual-structure for joint filtering.
\newblock In {\em ICCV},  3406--3414.

\bibitem[\protect\citeauthoryear{Tomasi and
  Manduchi}{1998}]{tomasi1998bilateral}
Tomasi, C., and Manduchi, R.
\newblock 1998.
\newblock Bilateral filtering for gray and color images.
\newblock In {\em ICCV},  839--846.
\newblock IEEE.

\bibitem[\protect\citeauthoryear{Wang \bgroup et al\mbox.\egroup
  }{2008}]{wang2008new}
Wang, Y.; Yang, J.; Yin, W.; and Zhang, Y.
\newblock 2008.
\newblock A new alternating minimization algorithm for total variation image
  reconstruction.
\newblock {\em SIAM Journal on Imaging Sciences} 1(3):248--272.

\bibitem[\protect\citeauthoryear{Wang, Wong, and
  Heng}{2006}]{wang2006deringing}
Wang, G.; Wong, T.-T.; and Heng, P.-A.
\newblock 2006.
\newblock Deringing cartoons by image analogies.
\newblock {\em ToG} 25(4):1360--1379.

\bibitem[\protect\citeauthoryear{Xu \bgroup et al\mbox.\egroup
  }{2011}]{xu2011image}
Xu, L.; Lu, C.; Xu, Y.; and Jia, J.
\newblock 2011.
\newblock Image smoothing via l 0 gradient minimization.
\newblock In {\em ToG}, volume~30,  174.
\newblock ACM.

\bibitem[\protect\citeauthoryear{Xu \bgroup et al\mbox.\egroup
  }{2012}]{xu2012structure}
Xu, L.; Yan, Q.; Xia, Y.; and Jia, J.
\newblock 2012.
\newblock Structure extraction from texture via relative total variation.
\newblock {\em ToG} 31(6):139.

\bibitem[\protect\citeauthoryear{Xu, Zheng, and Jia}{2013}]{xu2013unnatural}
Xu, L.; Zheng, S.; and Jia, J.
\newblock 2013.
\newblock Unnatural l0 sparse representation for natural image deblurring.
\newblock In {\em CVPR},  1107--1114.

\bibitem[\protect\citeauthoryear{Yang \bgroup et al\mbox.\egroup
  }{2014}]{yang2014color}
Yang, J.; Ye, X.; Li, K.; Hou, C.; and Wang, Y.
\newblock 2014.
\newblock Color-guided depth recovery from rgb-d data using an adaptive
  autoregressive model.
\newblock {\em TIP} 23(8):3443--3458.

\bibitem[\protect\citeauthoryear{Zhang \bgroup et al\mbox.\egroup
  }{2014}]{zhang2014rolling}
Zhang, Q.; Shen, X.; Xu, L.; and Jia, J.
\newblock 2014.
\newblock Rolling guidance filter.
\newblock In {\em ECCV},  815--830.
\newblock Springer.

\bibitem[\protect\citeauthoryear{Zhang \bgroup et al\mbox.\egroup
  }{2015}]{zhang2015segment}
Zhang, F.; Dai, L.; Xiang, S.; and Zhang, X.
\newblock 2015.
\newblock Segment graph based image filtering: fast structure-preserving
  smoothing.
\newblock In {\em ICCV},  361--369.

\bibitem[\protect\citeauthoryear{Zhang, Kwok, and
  Yeung}{2004}]{zhang2004surrogate}
Zhang, Z.; Kwok, J.~T.; and Yeung, D.-Y.
\newblock 2004.
\newblock Surrogate maximization/minimization algorithms for adaboost and the
  logistic regression model.
\newblock In {\em ICML},  117.
\newblock ACM.

\end{thebibliography}
}

\end{document}